\documentclass[twoside,english]{elsarticle}
\usepackage[utf8]{inputenc}
\usepackage[T1]{fontenc}

\usepackage{graphicx}
\usepackage{dcolumn}
\usepackage{bm}
\usepackage[normalem]{ulem}
\usepackage{stmaryrd}
\usepackage{graphics}
\usepackage{amsmath}
\usepackage{amssymb}
\usepackage{amsfonts}
\usepackage{mathrsfs}
\usepackage{latexsym}
\usepackage[bbgreekl]{mathbbol}
\usepackage{tikz}
\usepackage{todonotes}
\usepackage{hyperref}
\hypersetup{colorlinks=true, citecolor=blue, urlcolor=blue, linkcolor=blue}
\usepackage{color}

\usepackage[utf8]{inputenc}

\newcommand\Overline[2][1pt]{%
    \begin{tikzpicture}[baseline=(a.base)]
      \node[inner xsep=0pt,inner ysep=1.5pt] (a) {$#2$};
      \draw[line width= #1] (a.north west) -- (a.north east);
    \end{tikzpicture}}
\newcommand\bbar[1]{\Overline[0.8pt]{#1}}

\newcommand{\circledstar}{%
  \mathbin{%
    \tikz[baseline=(star.base)]{
      \node[shape=circle, draw, line width=0.4pt, minimum size=1ex, inner sep=0pt] (star) 
      {\raisebox{0.1ex}{\scalebox{0.9}{$\star$}}};
    }%
  }%
}

\newcommand{\cB}{\mathcal{B}}

\newcommand{\cT}{\mathcal{T}}

\newcommand{\cP}{\mathcal{P}}
\newcommand{\cD}{\mathcal{D}}
\newcommand{\cL}{\mathcal{L}}
\newcommand{\cE}{\mathcal{E}}
\newcommand{\bs}{\boldsymbol{s}}

\newcommand{\g}{\boldsymbol{g}}
\newcommand{\0}{\boldsymbol{0}}
\newcommand{\e}{\varepsilon}
\newcommand{\be}{\boldsymbol{e}}
\newcommand{\ba}{\boldsymbol{a}}
\newcommand{\bb}{\boldsymbol{b}}
\newcommand{\bc}{\boldsymbol{c}}

\newcommand{\bk}{\boldsymbol{k}}

\newcommand{\bx}{\boldsymbol{x}}
\newcommand{\by}{\boldsymbol{y}}
\newcommand{\bn}{\boldsymbol{n}}

\newcommand{\bu}{\boldsymbol{u}}

\newcommand{\cF}{\mathcal{F}}
\newcommand{\cG}{\mathcal{G}}
\newcommand{\V}{\textbf{V}} 
\newcommand{\B}{\textbf{B}} 
\newcommand{\G}{\textbf{G}}  
\renewcommand{\P}{\textbf{P}} 
\newcommand{\I}{\textbf{I}} 
\newcommand{\U}{\textbf{U}} 
\newcommand{\A}{\textbf{A}}

\renewcommand{\bbeta}{\boldsymbol{\beta}}

\newcommand{\bet}{\boldsymbol{\eta}}
\newcommand{\bxi}{\boldsymbol{\xi}}
\newcommand{\bsigma}{\boldsymbol{\sigma}}

\newcommand{\bomega}{\boldsymbol{\omega}}

\newcommand{\beps}{\boldsymbol{\varepsilon}}
\newcommand{\bpsi}{\boldsymbol{\psi}}

\newcommand{\btau}{\boldsymbol{\tau}}

\newcommand{\Div}{\text{Div}}
\newcommand{\Curln}{\text{Curl}_n}

\newcommand{\D}{\Delta}
\newcommand{\tr}{\text{tr}\hspace{0.5pt}}

\newcommand{\sym}{\text{sym}\hspace{0.5pt}}

\newcommand{\Curl}{\text{Curl}\hspace{0.5pt}}

\newcommand\beq{\begin{equation}}
\newcommand\beqn{\begin{eqnarray}}
\newcommand\eeqn{\end{eqnarray}}
\newcommand\eeq{\end{equation}}
\newcommand{\trsp}{\hspace{-1pt}^{\textsf{T}}\hspace{-2pt}}

\newcommand\jump[1]{\llbracket{#1}\rrbracket}

\newcommand\Vtau{\V_{\hspace{-1pt}\tau}}
\newcommand{\hodge}{\hspace{-1pt}\star}

\DeclareFontShape{OMX}{cmex}{m}{b}{<-> cmexb10}{}
\SetSymbolFont{largesymbols}{bold}{OMX}{cmex}{m}{b}
\usepackage{amsmath,bm}

\begin{document}

\begin{frontmatter}

\title{Micro-displacement tensor}

\date{}

\author[Galway]{G.~Zurlo\corref{cor1}}

\ead{giuseppe.zurlo@universityofgalway.ie}

\cortext[cor1]{Corresponding author}

\author[ESPCI]{L.~Truskinovsky }


\address[Galway]{School of Mathematical and Statistical Sciences, University of Galway, University Road, Galway, Ireland}

\address[ESPCI]{PMMH-ESPCI, 7 Quai Saint-Bernard, 75005 Paris, France}

\begin{abstract}
We propose an extended kinematics  of nominally elastic continuum solids allowing one to describe their mechanical interaction with micro-scale loading devices. The main new ingredient is the concept of a micro-displacement tensor which extends the conventional description of the deforming elastic solids in terms of macroscopic displacement vectors. We show that micro-displacement tensors are particularly useful in dealing with active incompatibility acquisition and its subsequent passive relaxation. We use the proposed approach to describe the energetics of surface deposition while accounting for the presence of micro-mechanical controls.To illustrate the effectiveness of the new conceptual scheme we present two case studies: crystallization from a melt resulting in pre-stress, and winding of a coil with controlled pre-stretch.  
\end{abstract}


\end{frontmatter}{}


\section{Introduction}

Recent   attempts to mimic biological prototypes in technological applications highlight the necessity  to extend the conventional kinematic framework of solid mechanics. One of the most important challenges of this type is to learn how   to deal in continuum framework with active materials capable of acquiring  and manipulating  mechanical ``information''. The latter can be  stored  in the form of geometric incompatibility    \cite{Han-Nature-2008, Moshe-2018, Gabrielov-pnas-1996}. Since such incompatibility    can be  implanted  into a solid material, the question arises  how this  can be performed in a controlled way. A closely related  question is how the stored ``information'' of this type can be  either  erased or  expanded. 
  
It is  well known that  the usual  defects   of crystal lattices representing sources of incompatibility can be     generated   using  conventional macroscopic loads.  However,   the control of the  distribution  of such defects is rather limited. An alternative approach, which we pursue in this paper,  would be to incorporate   incompatibility   through the interaction  of a solid with external micro-scale ``agents''.  The latter can be either integrated into the bulk or act  on the  surface \cite{Ziepke-nature-22, Baker-naturecomm-19}.  A technologically relevant example of a surface related   micro-mechanical activity is the interaction of a feeding head of a 3D printer with a deposited  elastic material \cite{Renzi-ejsma-24,Zaza-ejmsa-21} and this will be the main testing ground for the proposed general theory. In this regard the present  paper can be viewed as an attempt to  generalize  the ideas developed in our previous work dedicated to the design of 3D printing algorithms \cite{ZTPRL,ZTMRC,TZPRE}.
 
  As already intuitively clear, an uncontrolled   surface deposition/insertion   of  a new material can compromise the compatibility of an emerging elastic solid. It can then become  a source of  anelasticity  manifesting itself through the accumulation of residual stresses \cite{Eckart1948}. The problem of tempering   such  stresses comes to the forefront  when it is necessary   to  ensure that  the manufactured  material  responds in specific way to external loads. The corresponding control problem is fundamentally nonlocal, as the final configuration of  incompatibility depends on the whole history of the deposition/insertion  process. 
 
The processes of surface growth  are not only biologically relevant \cite{Wang-review-bioact-20, space-nature-2022}, but are also   technologically important, as in roll/coil winding and concrete deposition in structures.  In fact, they are  ubiquitous in nature, as in growth of crystals and gravitational accretion, and therefore the attempts of quantitative mechanical modeling of surface deposition  have a long history starting with the pioneering works \cite{Southwell, Trincher, Palmov1967, BrownGoodman, KingFletcher, Rashba}, further elaborated in \cite{Arutyunyan77, Metlov, Naumov88, Naumov94, Manzhirov89, Zabaras95, Skalak97, Drozdov98}. More recently, the process of surface deposition/ablation was modeled in the framework of the theory of nonlinear-elastic phase transitions with the non-trivial reference metric serving as a characterization of the reference state of the deposited ``phase'', e.g. \cite{Gambarotta2012, Ciarletta2013, Tomassetti2016, Gupta2018, Abi-Akl2019, Naghibzadeh2021, YaPr22,  YavariSafa2024,  Arroyo2024,Riccobelli2024}. In such   approaches, the advancement of the growth surface was either prescribed, say through the controlled rate of heat removal,  or regulated  by the kinetics of the ``effective'' phase transformation.  

Here we develop a new model where both the growth rate on the deposition surface and the evolution of the  growth-induced incompatibility, emerge as an outcome of the micro-activity of sub-continuum forces. To this end we  develop a novel  kinematic description of incompatible (Non-Eucleadian) solids allowing one to use continuum mechanical language while dealing with micro-deformations. The  main new ingredient of the proposed approach  is a non-classical micro-displacement \emph{tensor} which is  absent in   the conventional kinematic description of elastic solids focused exclusively on displacement \emph{vectors}.  In the developed framework, generalized  forces -- representing the mechanical activity  of the micro-scale ``agents'' --   emerge naturally as an outcome of  a thermodynamically consistent implementation of the extended kinematics of continuum. Since the   corresponding  micro-displacements are represented by   tensors,  it is  natural  that the conjugate micro-forces   also  emerge as  tensorial variables.  

To illustrate the developed theory, we present several detailed case studies highlighting different aspects of the proposed  quantitative description of micromechanical interactions. Our first example  concerns   the   stability of an interface separating a non-hydrostatically stressed solid from its hydrostatically loaded  melt and viewed in the perspective of surface deposition. While this problem was previously examined in an incompatibility-free setting, our study focuses on the peculiar features of crystallization under controlled pre-stress, which lead to the accumulation of incompatibility and the emergence of residual stresses in the newly formed solid. Our second case study concerns the process of  winding of a coil on a solid mandrel with a controlled pre-stretch of the tape. Using this example  we illustrate the idea that the acquired inelastic strain depends on the whole time dependent deposition protocol. Our last case study  deals  with  the peculiarities of the viscous relaxation  in the bulk  of the   inelastic strains accumulated during surface deposition. The presented examples   allow us to clarify the physical meaning behind the idea of  micro-controls and to discuss   their actual implementation in realistic situations. We note that the possibility of using the implied micro-scopic  ``agents''  in technological applications opens a new avenue in the design of   smart materials and structures, and can be expected to have significant impact on a broad range of engineering disciplines from construction to robotics.

The paper is organized as follows. In Section 2 we use Helmholtz decomposition of  elastic distortion to extend the classical kinematic description of macroscopic continuum elastic solids and to introduce the idea of a micro-scale displacement tensor. We then specify the structure of this tensor in the special case of materials with internal  layering and present  a simplified model of such layering allowing one to bridge microscopic and macroscopic descriptions. In Section 3  the developed extended kinematics is used to address the phenomenon of surface growth. To supplement  the purely mechanical description of the underlying activity of the microscopic  active ``agents''   we also present   the energetic and thermodynamic closure of the corresponding system of equations allowing one to describe self-consistently   the evolution of the growth surface. In Section 4 we illustrate the developed theory by addressing    the problem of solidification in a simple geometry. A  problem of coil winding, also  characterized by   simple geometry, is considered in Section 5. In Section 6 we suppress deposition and instead deal with the relaxation of acquired incompatibility.  Finally, in Section 7 we summarize our results and discuss potential  extensions of the proposed approach.

\section{Kinematics of incompatible solids }

While remaining in the framework of   classical linear elasticity theory  we   present the symmetric linear elastic strain tensor in the form
\beq\label{epsiloncomp}
\beps_e= \sym\nabla\bu
\eeq 
where $\bu=\bu(\bx)$ represents the  macroscopic displacement vector at a point $\bx$  in the   stress-free reference configuration; note that in \eqref{epsiloncomp}  we used the standard notation $$\sym\A=(\A + \A\trsp\,)/2$$ for the  symmetrization operator applied to an arbitrary second order tensor $\A$. The simple assumption \eqref{epsiloncomp} needs to be modified in the case of non-compatible deformations, where instead of \eqref{epsiloncomp} we  write 
\beq \label{epsiloncomp1}
\beps_e= \sym\nabla\bu-\beps_p 
\eeq
where  the  tensorial field  $\beps_p(\bx)$ represents  an   inelastic (plastic) strain.This implies that this field is not a symmetrized gradient of any displacement field, hence the associated incompatibility tensor is non-trivial: \beq\label{curlcurlep1}
\bet = -\Curl\Curl\beps_p \neq \0. 
\eeq
While the  field $\beps_p(\bx)$ formally brings into the theory six functional degrees of freedom, three of them describe  the compatible part of inelastic deformation and can be potentially   absorbed into the vector field $\bu(\bx)$. The remaining  three degrees of freedom describe  the incompatibility tensor field $\bet (\bx)$, which is by definition symmetric and divergence free.

\subsection{Micro-displacements in the bulk}

To characterize the  response of incompatible solids whose  inelastic  strain  is  generated in the process of surface deposition,  we now specialize the expression for the  tensor field  $\beps_p$ by assuming that 
\beq\label{epsilongen}
\beps_e = \sym\left(\nabla\bu + \Curl\U\right)
\eeq
where $\U(\bx)$ is a new tensor field which  we interpret as describing   microscopic inelastic displacements. It is clear that, due  to its structure,  the term  containing the tensor field $\U$ cannot be fully reduced to a compatible vector field  $\bu$.  The implied representation 
\beq \label{epsilongen1}
\beps_p = -\sym\left(\Curl\U\right) 
\eeq
can be interpreted as an attempt to introduce a generalized displacement field responsible for the inelastic strain, which enables a richer kinematics relative to standard compatible elasticity. In what follows, we refer to $\U$ as the \textit{micro-displacement tensor}. 

An  irreducible representation of  the incompatible tensor fields  $\U(\bx)$, containing only three independent functional degrees of freedom,  was proposed  by Beltrami \cite{Beltrami,FosdickRoyer2005,Gu1973,VG2014}. However, for the purpose of describing surface deposition, this representation is overly general, therefore we restrict our attention to a pertinent subclass of micro-displacement tensor fields, still parameterized by three scalar functions.

\subsection{Micro-displacements at the boundary}

To motivate the adopted structure of the micro-displacement tensor $\U$ and to clarify its relation to the process of layered deposition, consider a body $\cB$ with boundary $\partial\cB$. Introducing the Cauchy stress tensor $\bsigma$, defined as the derivative of the elastic energy density $W(\beps_e)$ with respect to the elastic strain, $\bsigma = \partial_{\beps_e} W$,
we may then use \eqref{epsilongen} to express  
\beq\label{identityCurl}
\int_{\cB} \bsigma \cdot \beps_e = \int_{\cB} \left( -\bu \cdot \Div\bsigma + \U \cdot (\Curl\bsigma)\trsp\, \right) + \int_{\partial\cB} \cT_{\text{bnd}}
\eeq
where 
\beq\label{bndgen}
\cT_{\text{bnd}} = \bsigma \bn \cdot \bu + \bsigma \cdot \U(\hodge\bn)\trsp.
\eeq
Here the notation $\hodge\bn$ is used for the skew-symmetric tensor with axial vector $\bn$, defined component-wise by the formula
\beq
(\hodge\bn)_{ij} = \varepsilon_{ikj} n_k
\eeq
where $\varepsilon_{ikj}$ is the Levi-Civita symbol\footnote{For arbitrary vectors $\ba,\bb$, the tensor $\hodge\ba$ acts on $\bb$ as $(\hodge\ba)\bb = \ba \times \bb$, and it satisfies the identity $(\hodge\ba)^2 = -\P_a$, where $\P_a=\I-\ba\otimes\ba$ is the orthogonal projection onto the plane normal to $\ba$. For the sake of notation, we will always intend that the operator $\hodge$ always acts on the vector at its right side, therefore if for example $\A$ is a tensor and $\ba$ a vector,  by $\A\hodge\ba$ we will intend the composition of a tensor $\A$ with the skew-symmetric tensor $\hodge\ba$, that is $(\A\hodge\ba)_{ij}=A_{i\ell}(\hodge\ba)_{\ell j}=A_{i\ell} \varepsilon_{\ell kj} a_k$.}. 

The representation of the boundary term \eqref{bndgen} suggests  that the introduction of the micro-displacement tensor $\U$ enables a richer set of boundary controls compared to standard elasticity.  To be more specific, we   split the full stress tensor $\bsigma$  on the boundary  into normal and tangential parts using the definitions
\beq \label{notations2}
\bs_n:=\bsigma\bn,
\qquad
\bsigma_{\tau} := \P_n \bsigma \P_n,  
\eeq 
where 
$$\P_n:=\I-\bn\otimes\bn$$
is the standard surface projection tensor. Note first that  $\bs_n$ is the classical traction vector, which performs work against macroscopic displacements $\bu$. To  specify the displacements against which the work is performed by the surface (tangential) stress tensor $\bsigma_{\tau}$, we use for the micro-displacement tensor at the boundary $\partial\cB$ the  ansatz
\beq\label{ansatzU1}
\U = \Vtau\hodge\bn 
\eeq
where the tensor $\Vtau$ is symmetric and tangential, i.e., $\Vtau = \Vtau\trsp$ and $\Vtau\bn = \0$. Then, 
 \beq
\cT_{\text{bnd}} = \bs_n\cdot \bu + \bsigma_{\tau} \cdot \Vtau.
\eeq
Accordingly, the ansatz \eqref{ansatzU1} allows one to set apart  the  contributions of the normal  and tangential components of stress at the boundary.

\subsection{Layering ansatz}

Suppose next that we deal  with solid bodies emerging as a result of sequential surface deposition of material layers as in the case of   3D printing.  Physically, one can think about solid bodies emerging as a result of sequential surface deposition of material layers, which may be pre-stressed independently. If such body ``remembers'' the micro-deposition through the associated ``frozen''  local layering, characterized by the prescribed vector field $\bn(\bx)$, we may assume that the parametric representation of $\U$ given by  \eqref{ansatzU1} holds everywhere in the bulk of the body, so that 
\beq\label{ansatzU}
\U(\bx)  =  \Vtau(\bx)\hodge\bn(\bx) \qquad \bx\in\cB
\eeq
where $\V_\tau(\bx)$ is a symmetric  tensor field which satisfies  the constraints 
\beq\label{ansatzU3}
\Vtau(\bx)\bn(\bx)=\0,\qquad\Vtau(\bx)=\Vtau(\bx)\trsp, \qquad \bx\in\cB. 
\eeq
Because the field $\V_\tau$ is both tangential and symmetric, it can be expressed in a local basis represented by the normal vector field $\bn$ and two orthogonal vector fields  $\be_1, \be_2$ that span the plane orthogonal to $\bn$:  
\beq\label{Vtau}
\V_\tau=
\left(
\begin{array}{ccc}
v_{11} & v_{12} & 0 \\
v_{12} & v_{22} & 0 \\
0 & 0 & 0
\end{array}
\right) 
\eeq
where  the three independent parameters are $v_{11}(\bx)$, $v_{12}(\bx)$ and $v_{22}(\bx)$. We can then conclude  that the field $\U$ constrained  through \eqref{ansatzU}   carries exactly  three independent functional degrees of freedom.  Moreover, given that 
\beq\label{curlcurlep}
\bet = -\Curl\Curl\left(\sym\left(\Curl(\Vtau \hodge\bn)\right)\right )
\eeq
the ansatz \eqref{ansatzU} can be thought as representing an incompatibility tensor \( \bet \)  of a generic type since the latter is also  symmetric and  divergence-free and therefore also carrying  exactly three independent functional degrees of freedom.

To further elucidate the physical meaning of  the tensor field $\Vtau(\bx)$, we  observe that when \eqref{ansatzU} holds, the average inelastic  strain in the body can be written as 
\beq\label{avplast1}
\bbar{\beps}_p = \frac{1}{\text{vol}(\cB)}\int_{\cB} {\beps}_p = - \frac{1}{\text{vol}(\cB)}\int_{\partial\cB}\Vtau.
\eeq
We can then  write the expression for the average elastic strain in the form
\beq\label{avstrain}
\bbar\beps_e=\frac{1}{\text{vol}(\cB)}\int_{\partial\cB} \left(\bu\odot\bn + \Vtau\right)
\eeq
where we used the standard definition of the 
 symmetrized tensor product:  $$\ba\odot\bb=\sym(\ba\otimes\bb)$$  for arbitrary vectors $\ba$ and vector $\bb$. To emphasize that  $\Vtau$ in \eqref{avstrain} depends on $\bn$, we can return to the description of micro-displacements in terms of the tensor $\U$ and use  its relevant projection
 $\U_\tau: =\P_n\U\P_n$  to recast \eqref{avstrain} in the form
\beq\label{avstrain1}
\bbar\beps_e=\frac{1}{\text{vol}(\cB)}\int_{\partial\cB} \left(\bu\odot\bn - \tilde\U_\tau\circledstar \bn\right), 
\eeq
where $\tilde\U_\tau:=\U_\tau-\tfrac{1}{2}(\tr\U_{\tau})\P_n$ is the deviator of $\U_\tau$; note that in \eqref{avstrain1}   we introduced a  new notation $$\A\circledstar\ba=\sym(\A\hodge\ba)$$ for arbitrary tensor $\A$ and vector $\ba$. 
The symmetric nature of the relation \eqref{avstrain1} serves as a reminder  that the macro-displacement vector field $\bu(\bx)$ and the micro-displacement tensor  field $\U(\bx)$ play  parallel and complementary roles in the kinematics of the  targeted  class  of inelastic (incompatible) solids.

\subsection{Microscopic perspective}

\begin{figure}[h!]
    \centering
    \includegraphics[width=0.9\linewidth]{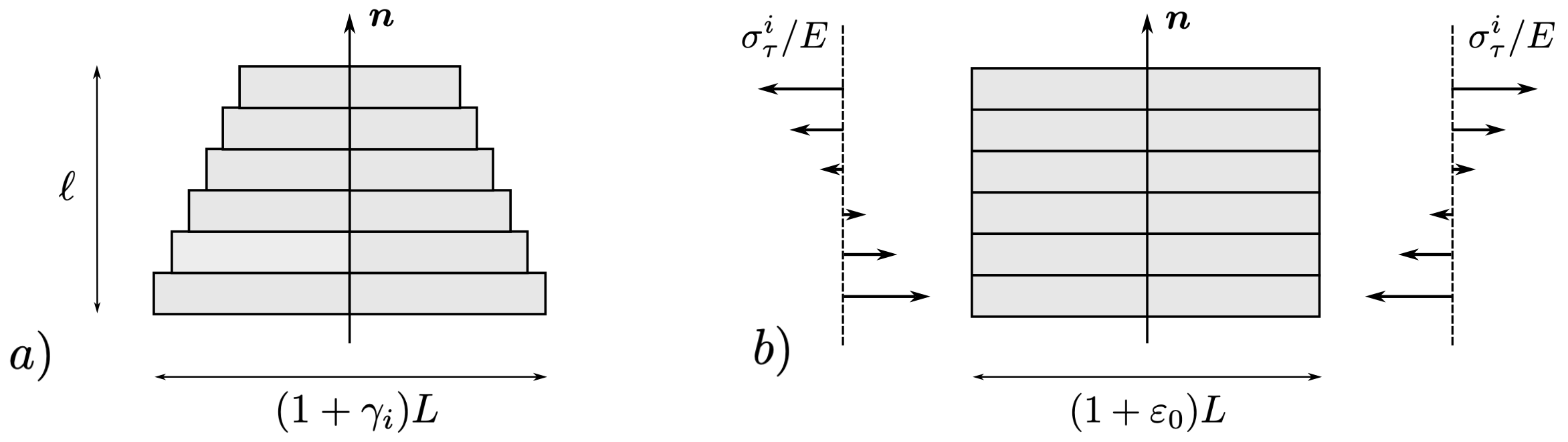}
    \caption{{\label{Vstack} Residual stresses in an incompatible arrangement  of $m$ layers, oriented perpendicularly to the direction $\bn$. The layers, having initially uneven stress-free lengths, see $(a)$, are stretched or compressed to be of the same size  and then glued together, see  $(b)$.}}
\end{figure}

A microscopic interpretation of the tensor field $\V_\tau(\bx)$ is obtained by considering the simplest one-dimensional micro-layering, represented by a stack of thickness $\ell$   made of  $m$ flat elastic layers, each  of thickness $\ell/m$. As shown in Fig.\ref{Vstack}, the layers are  piled up perpendicularly to a fixed direction $\bn$. 

Suppose  that the layers have different lengths in the reference state, see Fig.\ref{Vstack}(a) and therefore have to be  pre-stressed in order to fit the stack with even length, see Fig.\ref{Vstack}(b). Denote by $\beps_p^i=\gamma_i\P_n$, with $i=1,..,m$, the corresponding  residual pre-strains  which are assumed to be   known. To be consolidated into a single stack with uniform length perpendicularly to $\bn$, see Fig.\ref{Vstack}(b), each  layer has to be  deformed elastically by corresponding lateral  tractions $\sigma_{\tau}^i$, see Fig.\ref{Vstack}(c). If we assume   that the deformation is strictly lateral and we adopt again the framework of classical linear elasticity,   we can express the required lateral stress in each layer in the form 
\beq\label{sigmai}
\sigma_{\tau}^i=E(\e_0 -\gamma_i)
\eeq
where $E$ is an elasticity modulus and $\e_0 $ is  the total strain  in the assembled stack which, according to our assumption, should be the same  for  all layers. As the layers are brought to the same horizontal length, they undergo relative sliding which compensates for the original inhomogeneous  uni-axial pre-strain. 

Suppose next that  the consolidated  layers are bonded by a perfect ``glue'' which prevents subsequent relative sliding. Then the   constraints are removed and the  resulting composite body is  allowed to relax    and  reach elastic equilibrium. Suppose further that during such  relaxation the deformation in each layer is constrained to remain uniaxial, so that no bending can occur. 

Due to the presence of the glue, such relaxation must be necessarily homogeneous, with the stack expanding or contracting in the lateral direction. The resulting equilibrated structure will  carry residual stresses while not being loaded externally, which means that the composed body will contain embedded distributed incompatibility. 

To characterize equilibrium we use the fact  that in the relaxed state the resultant force on the lateral edges must be equal to zero,  cf. \cite{Palmov1967}: 
\beq \label{epsavdiscr1}
\sum_1^m\sigma_{\tau}^i=0.
\eeq
If we substitute  \eqref{sigmai} into this equation we obtain that 
 \beq\label{epsavdiscr}
\e_0 \equiv \langle \gamma \rangle:= \frac{1}{m} \sum_1^m\gamma_i
\eeq
where $\langle \gamma \rangle$ is the average inelastic strain in the stack.  Therefore, the stress in the stack is
\beq\label{resstack}
\sigma_{\tau}^i=E(\langle \gamma \rangle -\gamma_i). 
\eeq
This expression reveals that the the stress in such composite body vanishes if and only if the plastic strains $\gamma_i$ are uniform. As we now show, this corresponds precisely to the case where  the  plastic strain becomes  compatible and the micro-displacement tensor becomes trivial.

To compute the corresponding micro-displacement tensor, note  that the inelastic strain in the stack may be written in the form 
\beq
\label{betapstack}
 \beps_p=\sum_{i=1}^m\gamma_i\chi_i(\bx)\P_n
\eeq
where $\chi_i(\bx)=H(z-z_{i-1}) - H(z-z_i)$ are the characteristic functions which distinguish  individual  layers, $H(z)$ is the Heaviside function, $z=\bx\cdot\bn$ is the coordinate perpendicular to the direction of pre-stressing and  $z_i=i\ell/m$. If we now rewrite  \eqref{betapstack} using the decomposition  $\beps_p = - \sym\left(\nabla\bu  + \Curl\U \right)$, 
with $\U(\bx) =\Vtau(\bx)\hodge\bn(\bx)$, we obtain  that the  (compatible) macro displacements are of the form 
\beq
\bu(\bx)= - \gamma_1\P_n \bx
\eeq 
whereas the incompatible (micro) displacement tensor is
\beq\label{Vdiscrete}
\Vtau(\bx)= - \P_n \sum_{i=1}^{m-1}\jump{\gamma_i} R(z-z_i). 
\eeq  
In this expression $R(z)=z H(z)$ is the ramp function, and $\jump{\gamma_i}=\gamma_{i+1}-\gamma_i$ are the jumps of strains between neighboring layers. 

As noted already from \eqref{resstack}, residual stresses vanish when all $\jump{\gamma_i}=0$ throughout the stack, meaning that the jumps $\jump{\gamma_i}$ are a direct measure of incompatibility under the assumed kinematic constraints, and they also explicitly appear in the micro-displacement tensor describing the incompatible displacements.

\subsection{Micro-displacement tensor and second order structured deformations}

Given that in the  discrete   model  discussed  in the previous subsection,  the micro-displacement tensor $\U (\bx)$ emerges due to discontinuous  strains in the  neighboring layers, it is  natural to link it to the nonclassical macro-variables introduced in the theory of second-order structured deformations \cite{OwenParoni2000}. 

We begin with the observation that strain discontinuities serve in general as sources for  micro-displacement tensors $\U$. Consider for instance a finite body textured by a sequence of infinitesimally thin internal layers $\cL_i$ of thickness $h = \ell/m$, where $\ell$ is the fixed characteristic length of the body. Assume that each layer undergoes a compatible distortion $\nabla \tilde{\bu}_i$, which may be discontinuous across the jump surface  $\mathcal{J}(\nabla \tilde{\bu}_i)$ separating $\cL_i$ and $\cL_{i+1}$, cf. \cite{Paroni}. In the   corresponding continuum description we can introduce the plastic strain
 \beq \label{epsilongen112}
\beps_p = -\sym(\bbeta_p). 
\eeq
Here the inelastic  distortion 
\begin{equation} \label{epsilongen11}
\bbeta_p = - (\Curl \U)\trsp
\end{equation}
 should be understood as the  limit
\begin{equation}\label{betapp}
\bbeta_p = \lim_{m\to\infty} \sum_{i=1}^m (\nabla \tilde{\bu}_i) \, \chi(\cL_i)
\end{equation}
where $\chi(\cL_i)$ is the characteristic function of layer $\cL_i$. The corresponding Nye tensor \cite{Nye53} takes the form
\begin{equation}\label{USD0}
\G = \Curl \bbeta_p = \lim_{m\to\infty} \sum_{i=1}^m (\hodge \bn_i) \jump{\nabla \tilde{\bu}_i}\trsp \delta_i
\end{equation}
where $\delta_i$ is the two-dimensional Dirac measure supported on the jump surface $\mathcal{J}(\nabla \tilde{\bu}_i)$, and $\bn_i$ is its unit normal. 

The representation \eqref{epsilongen11} suggests that 
\begin{equation}\label{PoissonU}
\D \U = \G\trsp 
\end{equation}
and therefore \cite{Jackson1998}
\begin{equation} \label{PoissonU1}
\U(\bx) = -\frac{1}{4\pi} \int_{\mathbb{R}^3} \frac{1}{\varrho(\bx, \by)} \G\trsp(\by)\, dV(\by)
\end{equation}
where  we considered for simplicity an unbounded domain and used the notations
$\varrho(\bx, \by) = \|\bx - \by\|$ for  the Euclidean distance between points $\bx$ and $\by$  and  $dV(\by)$ for  the volume element centered at $\by$. Substituting \eqref{USD0} into  \eqref{PoissonU1}, we conclude that away from the boundaries of the body 
\begin{equation}\label{USD}
\U(\bx) = \lim_{m\to\infty} \sum_{i=1}^m \frac{1}{4\pi} \int_{\cB \cap \mathcal{J}(\nabla \tilde{\bu}_i)} \frac{1}{\varrho(\bx, \by)} \jump{\nabla \tilde{\bu}_i(\by)} \hodge \bn_i(\by)\, dA_i(\by).
\end{equation}
Formulas similar to \eqref{USD}, but involving an additional ``localization'', are employed in the theory of second-order structured deformations to define macroscopic observables that account for microscale “corrections” to the conventional second gradients of the macroscopic fields $\bu(\bx)$, see, e.g., \cite{OwenParoni2000}. However, these corrections remain local. In contrast, in our theory, the corresponding macroscopic observables aim to “correct” the macroscopic displacement field $\bu(\bx)$ in a strongly nonlocal manner. Nevertheless, in both approaches, the jump discontinuities of the deformation gradient play a central role, giving rise to novel kinematical descriptors that emerge through the associated averaging, with or without concurrent localization.

\section{Surface deposition}

To illustrate the effectiveness of the proposed kinematic  extension of the conventional elastic continuum  model, we now turn to surface deposition. In particular, since this phenomenon involves continuous addition of prestressed layers of matter to the external boundary of a body,  it is a natural  playground for checking the validity of our ansatz for the micro-displacement tensor.

Observe first that the process of surface deposition implies the existence of an evolving reference domain $\cB_t$, where $t$ is time. Suppose that the new layers of material are continuously deposited on the external boundary of this domain $\Omega_t=\partial\cB_t$.  The latter may be defined by the corresponding surface coordinates $\bxi = (\xi_1,\xi_2)$ and by the mapping  
\beq
\bx=\bpsi(\bxi,t).
 \eeq
The function  $\bpsi(\bxi,t)$ can be always chosen in such a way that the vector  $\dot{\bpsi}(\bxi,t)$ is normal to $\Omega_t$, where the dot denotes differentiation with respect to time. Thus we can then write 
\beq
\dot{\bpsi}=D\bn
\eeq 
where $\bn$ is the normal to the surface $\Omega_t$ and $D=||\dot{\bpsi}||$ is the normal velocity of $\Omega_t$. Equivalently, the growth surface may  be represented as the level-set of a function 
\beq
\vartheta(\bx)=t
\eeq
such that $\vartheta(\bpsi(\bxi,t))=t$ for all $\bxi$. We can then write
\beq \label{n1}
\bn(\bx)=D\nabla\vartheta
\eeq
and therefore
\beq \label{D}
D(\bx)=||\nabla\vartheta||^{-1}. 
\eeq 
Note that according to \eqref{n1}  the local orientation of the layers past their acquisition 
remains  fixed, which means that the vector field $\bn=\bn(\bx)$ is effectively ``frozen''. 
 \begin{figure}[h]
    \centering
    \includegraphics[width=0.5\linewidth]{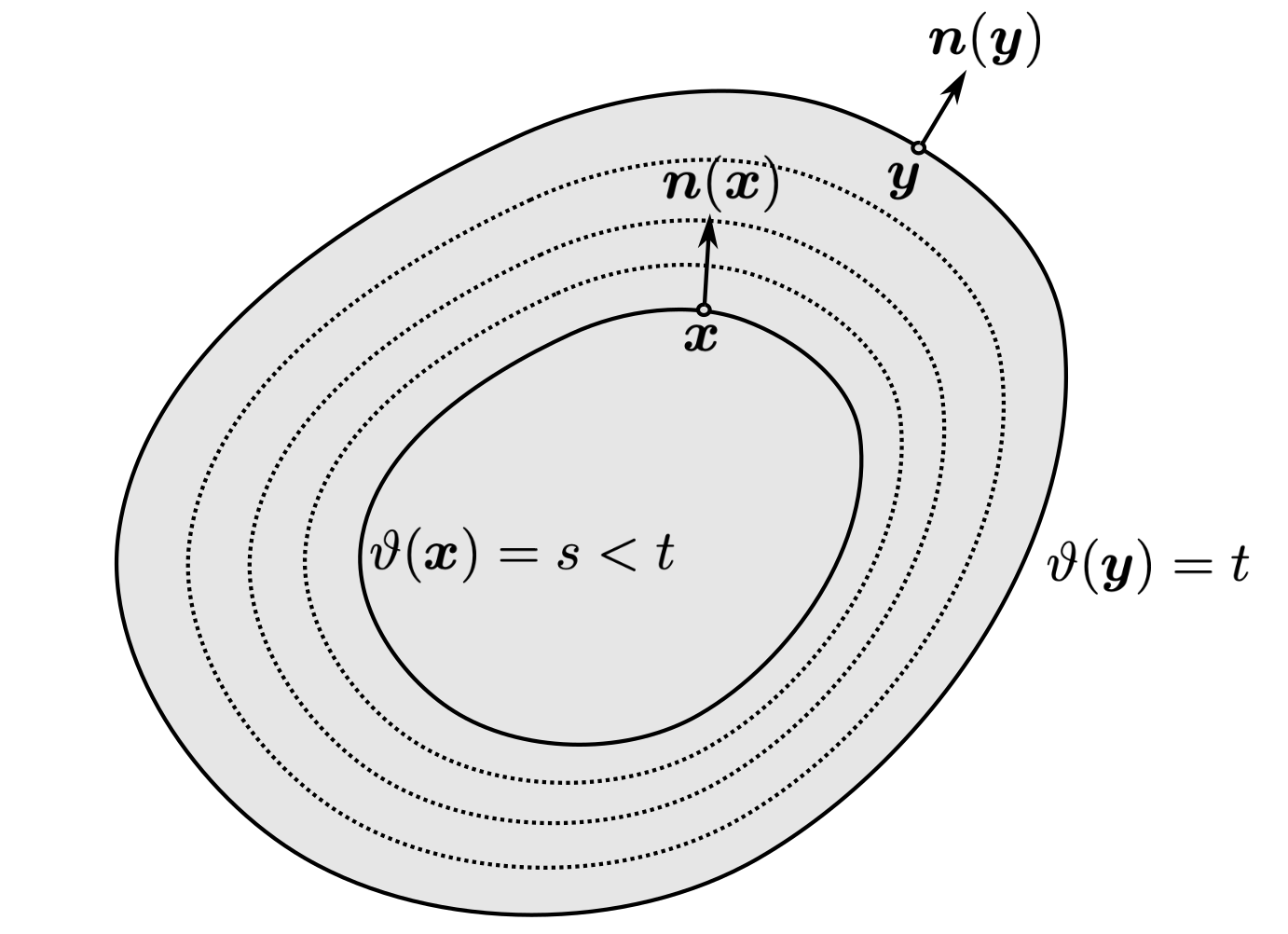}
    \caption{{\label{onion} The growing body which retains a  memory of its internal layering emerging during the deposition process. The local orientation of the layer $\bn(\bx)$ is determined at the moment when the growth surface passes through the point $\bx$ and remains unchanged thereafter.
   }}
\end{figure}

 \subsection{Dynamics  \label{dynamics1}}  

To formulate   thermodynamically consistent dynamic   theory of surface growth, we can  use the conventional machinery of continuum mechanics. We assume that the material model  is fully characterized by the specific free energy density $W(\beps_e)$ which is a function of the linear elastic strain tensor  $\beps_e(\bx,t)$ defined for $\bx\in\cB_t$ by the relation
\beq \label{ansatz2}
\beps_e=\sym\left(\nabla\bu + \Curl({\Vtau} \hodge\bn(\bx)) \right)
\eeq
where $\bu(\bx,t)$ is the evolving macro-displacement vector field and $\Vtau(\bx,t)$ is the evolving micro-displacement tensor tangential field,  satisfying
\beq\label{Vansatz}
\left\{
\begin{array}{lll}
\Vtau(\bx,t)=\Vtau\hspace{-4pt}\trsp(\bx,t)\\
\Vtau(\bx,t)\bn(\bx) = \0.
\end{array}
\right. 
\eeq
The next step is to compute the rate of change of the elastic energy 
$
\cE=\int_{\cB_t}W
$
while taking into account   the time dependence of  $\cB_t$. Using standard transformations we obtain
\beqn \label{e1}
\dot\cE &= 
&\int_{\cB_t}\left(-\dot\bu\cdot\Div\bsigma - \dot{\V}_{\tau}\cdot\Curln \bsigma\right) +\nonumber\\
&& \int_{\Omega_t}\left(\mathring\bu\cdot \bsigma\bn + \mathring{\V}_{\tau}\cdot\bsigma_{\tau}\right) + \nonumber\\
&& \int_{\Omega_t}D\left( W  -  \bsigma\bn\cdot \partial_n\bu  - \bsigma_{\tau}\cdot \partial_n{\Vtau} \right),
\eeqn 
where $\bsigma=\partial_{\beps_e}W$ is the   Cauchy stress while 
 $\bsigma_{\btau}:=\P_n\bsigma\P_n$ is the surface stress at the boundary. We denoted by $\partial_n$ the directional derivative along $\bn$, and  used   convective derivatives as in $\mathring\bu :=  \dot\bu  + D\,\partial_n\bu$ and 
$\mathring{\V}_{\hspace{-1pt}\tau} :=  \dot{\V}_{\hspace{-1pt}\tau}  + D\, \partial_n{\Vtau}$. In addition, we introduced a new notation  
\beq\label{Curln}
\Curln \bsigma=-\P_n\sym\left((\hodge\bn)\Curl\bsigma\right)\P_n.
\eeq
Assume next that the external power can be written in the form:
\beq\label{power}
\cP = \int_{\Omega_t}
\left(\bs_n\cdot\mathring\bu + {\bsigma}_a\cdot\mathring{\V}_{\hspace{-1pt}\tau}+ W_a D
\right)+ \int_{\cB_t}
\left(\bb\cdot\dot\bu +  \B_a\cdot\dot{\V}_{\hspace{-1pt}\tau}\right). 
\eeq
Here $\bs_n$ is the classical surface traction while $\bb$ introduces classical bulk force. The last term in the surface integral in \eqref{power} ( proportional to $D$) is   also classical, with the constant $W_a$ representing the  reference energy of the deposited  material.

The  remaining  terms in \eqref{power}, proportional to $\dot{\V}_{\hspace{-1pt}\tau}$ in the bulk and $\mathring{\V}_{\hspace{-1pt}\tau}$ on the surface $\Omega_t$, are new. They can be viewed as representing  the  microscopic ``activity''   involved in the creation  of incompatibility.  Specifically,   the incompatibility buildup on the deposition surface $\Omega_t$, is driven by   the tensorial ``generalized  forces''  ${\bsigma}_a$. Similarly, the incompatibility creation in the bulk    of $\cB_t$  is controlled by the   tensorial ``generalized  forces'' $\B_a$.  In view of the special structure of the tensor ${\Vtau}$, the conjugate tensorial fields $\B_a$ and $\bsigma_a$  are symmetric and tangential  so that that $\B_a=\B_a\trsp$, $\B_a\bn=\0$ and  $\bsigma_a=\bsigma_a\trsp$, $\bsigma_a\bn=\0$. 

We are now in a position to derive the dynamic equations of the model including the equation governing  the evolution of the   micro-displacement tensor.  For simplicity we ignore inertial effects and neglect the conventional  viscosity. We also assume that the system is isothermal. Then we can compute the  dissipation rate  
\beq\label{dissipation}
\begin{array}{lll}
\cP - \dot\cE=\cD & = &  \int_{\cB_t}\dot\bu\cdot\left(\bb+\Div\bsigma\right) \\
& + & \int_{\cB_t}\dot{\V}_{\hspace{-1pt}\tau}\cdot(\B_a+\Curln \bsigma)\\
& + & \int_{\Omega_t}\mathring\bu\cdot (\bs_n- \bsigma\bn)\\
& + & \int_{\Omega_t}\mathring{\V}_{\hspace{-1pt}\tau}\cdot ({\bsigma}_a-\bsigma_{\tau})\\
& + & \int_{\Omega_t}D\left({W_a} -  W  +  \bsigma\bn\cdot\partial_n\bu + \bsigma_{\tau}\cdot\partial_n{\Vtau}\right)\geq 0. 
\end{array}
\eeq
We assume  for simplicity  that the    mechanical response,  associated with the   macroscopic rates  $\dot\bu $ in the bulk of the domain $\cB_t$ and $\mathring\bu$ on the growth surface $\Omega_t$, is purely elastic and therefore dissipation free. We can then localize the corresponding terms in \eqref{dissipation} to obtain    the  field equations 
\beq\label{Pbulk1}
\cB_t:\quad 
\Div\,\bsigma + \bb = \0
\eeq
and the boundary conditions
\beq\label{Pbnd1}
\Omega_t:\quad 
\bsigma\bn = \bs_n. 
\eeq
For the microscopic displacements we assume  the simplest  overdamped relaxational dynamics.   Specifically we assume  that the dissipative potential  is an  isotropic   quadratic function of the non-classical rates $\dot{\V}_{\hspace{-1pt}\tau}$ in the bulk of the domain $\cB_t$  and   $\mathring{\V}_{\hspace{-1pt}\tau} $ on the growth surface $\Omega_t$. This gives  the kinetic equations in the bulk
\beq\label{Pbulk2}
\cB_t:\quad
\B_a+\Curln \bsigma=\alpha\dot{\V}_{\hspace{-1pt}\tau}
\eeq
that is supplemented by  the corresponding   boundary conditions 
\beq\label{Pbnd2}
\Omega_t:\quad
 {\bsigma}_a-\bsigma_{\tau} = \gamma \mathring{\V}_{\hspace{-1pt}\tau}.
\eeq
Note  that in \eqref{Pbulk2} and \eqref{Pbnd2} we introduced the phenomenological  coefficients  $\alpha \geq 0$ and $\gamma \geq 0$ characterizing the associated  volumetric and surface   relaxation rates, respectively. Finally, if we similarly assume that the deposition rate $\dot{\bpsi}$ is not prescribed and, instead, we associate with it a quadratic dissipative potential, we obtain  kinetic relation regulating the surface growth  process in the form 
\beq\label{eqD3}
\Omega_t:\quad
W_a -  W  + \bsigma\bn\cdot \partial_n\bu  + \bsigma_{\tau}\cdot \partial_n{\Vtau}   = \beta D
\eeq
where  $\beta \geq 0$  is another phenomenological  coefficient.

\subsection{Limiting regimes}
We now discuss some  physically relevant special  regimes emerging as we make  particular  assumptions regarding the relative values  of the (appropriately nondimensionalized)  phenomenological rate constants $\alpha$, $\beta$ and $\gamma$.
 
For instance,  in the limiting case  $\gamma=0$,  meaning physically that  the associated relaxation time  is   small  at the time scale of the evolution of the system,  we essentially assume   that the equilibration on the boundary is  infinitely  fast. Then  our  Eq. \eqref{Pbnd2} reduces to an equality
 \beq\label{Pbnd21}
\Omega_t:\quad
\bsigma_{\tau} = \bsigma_a.   
\eeq
Under this assumption, the tangential component of the stress tensor $\bsigma$ on the growth surface is tightly controlled by the externally imposed “generalized forces” ${\bsigma}_a$. Together with \eqref{Pbnd1}, this implies that the entire stress tensor is prescribed on the deposition surface. This constitutive choice was adopted, for instance, in our previous works \cite{ZTPRL,ZTMRC,TZPRE}. In the absence of experimental evidence suggesting otherwise, we also adopt \eqref{Pbnd21} as one of the boundary conditions in the present study.   

The opposite limit, $\gamma=\infty$, effectively suggests  that  
\beq
\Omega_t:\quad
\mathring{\V}_{\hspace{-1pt}\tau}=\0.
\eeq 
  In this case the micro-displacement tensor   would have to be externally controlled  on the boundary. Note that for finite values of $\gamma$, the evolution described by \eqref{Pbnd21} may account, for example, for slippage between the newly added layers and the existing body, see for instance a related  discussion in  \cite{Arroyo2024}. 

Another important limiting case is when  $\beta=0$ in \eqref{eqD3}, meaning that the growth surface is always in equilibrium as the attachment kinetics is formally  instantaneous. This would mean that instead of \eqref{eqD3} we should use 
\beq\label{eqD31}
\Omega_t:\quad
W_a -  W  + \bs_n\cdot \partial_n\bu  + {\bsigma}_a\cdot \partial_n{\Vtau}   = 0
\eeq
wherein wa used \eqref{Pbnd21}. Note that in addition to the conventional macroscopic elasto-static Eshelby force $ W  - \bs_n\cdot \partial_n\bu$ we have in \eqref{eqD31} a matching microscopic  (active)  contribution $W_a  + {\bsigma}_a\cdot \partial_n{\Vtau}$. 

In the opposite limiting regime $\beta=\infty$, when growth is fully suppressed, the boundary condition \eqref{eqD3} is replaced by the condition 
\beq\label{eqD32}
\Omega_t:\quad
D=0
\eeq  
meaning that no boundary deposition is taking place.
 
Yet another interesting limiting case is when $\alpha=0$ in \eqref{Pbulk2} and the bulk relaxation of the field ${\Vtau}$ is   instantaneous. In this case we obtain the bulk equilibrium equation 
\beq \label{f}
\cB_t:\quad
\Curln \bsigma + \B_a=\0
\eeq
suggesting that the “generalized force” $\B_a$ fully governs $\Curln \bsigma$, consistent with our other implicit assumption that the physical macroscopic force $\bb$ controls $\Div\bsigma$. In this context, the externally imposed tensor field $\B_a$ acts as a source of incompatibility, giving rise to the associated residual stresses.  In the context of our microscopic model, the field $\B_a$ can be interpreted as representing the externally applied system of micro-forces acting within the stack of parallel layers. It effectively serves as a “glue” that prevents relative sliding between layers, thereby preserving the jumps in reference strains across neighboring layers.
  
In the opposite limiting regime $\alpha=\infty$  we obtain from   \eqref{Pbulk2}  that 
\beq \label{f1}
\cB_t:\quad
  \dot{\V}_{\hspace{-1pt}\tau}(\bx,t) =\0  
\eeq
and therefore ${\Vtau}={\Vtau}(\bx)={\Vtau}(\bx,\vartheta(\bx))$. This means that the micro-displacement tensor does not evolve past the moment of deposition and therefore the incompatibility acquired at deposition is  effectively ``frozen''.

\subsection{Relaxation of incompatibility \label{relaxation1}}

As already indicated, the micro-displacement tensor field $\V_{\hspace{-1pt}\tau}(\bx,t)$ can contain, in principle, not only an incompatible part which is responsible for the emergence of a non-trivial incompatibility tensor field $\bet(\bx,t)$, but also a compatible part which can be, in principle, absorbed into the macro-displacement vector field $\bu(\bx,t)$. As we see from the examples considered later in the paper, this is indeed the case even under very restricted geometric assumptions. The apparent contradiction with our  assumption that the fields  $\V_{\hspace{-1pt}\tau}(\bx,t)$  and $\bu(\bx,t)$ are independent, would not pose any problem if the adopted relaxation model for  the field $\V_{\hspace{-1pt}\tau}(\bx,t)$ would concern only its incompatible component. While the general analysis of this problem is beyond the task of this paper, we can still provide some arguments in support of the conjecture that the dynamics of the incompatible part of $\V_{\hspace{-1pt}\tau}(\bx,t)$ is largely independent from the evolution of its compatible part.

For simplicity, we consider the  special case when the  external sources potentially driving the production of incompatibility are absent. Specifically, assume that the (passive) bulk vectorial forces and the (active) bulk tensorial  forces can be neglected, so that 
\beq\label{Pbulk111}
\cB_t:\quad  \bb = \0,\quad \quad
\B_a=0
\eeq
In this case,  the dynamic equations reduce to 
\beq\label{Pbulk11}
\cB_t:\quad 
\Div\,\bsigma  = \0, \quad \quad
 \Curln \bsigma=\alpha\dot{\V}_{\hspace{-1pt}\tau}
\eeq
where the rate of relaxation of $\V_{\hspace{-1pt}\tau}(\bx,t)$,  governed by \eqref{Pbulk11}$_2$,  is controlled by the parameter $\alpha$. The question is whether such relaxation impacts both compatible and incompatible components of $\V_{\hspace{-1pt}\tau}(\bx,t)$ or it actually concerns only the incompatible part of $\V_{\hspace{-1pt}\tau}(\bx,t)$, while leaving the corresponding compatible component intact. 

To at least partially answer this question it is necessary to obtain  a description of the  final post-relaxation state  described by the equations 
\beq\label{Pbulk1111}
\cB_t:\quad 
\Div\,\bsigma  = \0, \quad \quad
 \Curln \bsigma=\0. 
\eeq
Even this limited question cannot be addressed here in full generality, and instead we consider the simpler problem of characterizing solutions of the system
\beq\label{div-curl-equi}
\left\{
\begin{array}{lll}
\Div\bsigma = \0, \\
\sym(\Curl\bsigma) = \0. 
\end{array}
\right.
\eeq
One can see that   we  replaced the  special  operator $\Curl_n \bsigma$ relying on the existence of a ``frozen'' field $\bn(\bx)$ by  a more general one,   $\sym(\Curl\bsigma)$.  To justify this choice we note that in all  examples considered later in the paper,  the special choice of the deposition geometry (radial and planar) ensures that such  special and general  problems are fully equivalent. 

Observe first that for isotropic linear elastic materials, one can construct a class of nontrivial \emph{compatible} deformations with $\beps_e = \sym\nabla\bu$ which  solves equations \eqref{div-curl-equi}. Indeed, in this case, the system \eqref{div-curl-equi} takes the form 
\beq\label{div-curl-equi-3}
\left\{
\begin{array}{lll}
\mu\,\D\bu + (\mu+ \lambda)\,\nabla\Div\bu  = \0, \\
\sym(\nabla\Curl\bu) = \0
\end{array}
\right.
\eeq
where we introduced Lamé constants $\mu$ and $\lambda$.   Note next that     \eqref{div-curl-equi-3}$_1$ can be also rewritten as
\beq \label{curl1}
\nabla\Div\bu = \frac{\mu}{2\mu+\lambda}\,\Curl\Curl\bu.
\eeq
From \eqref{div-curl-equi-3}$_2$ we now infer that the gradient of $\Curl\bu$ is a skew-symmetric tensor field that can be represented in the form  \cite{Gurtin-ICM} 
\beq\label{curlu}
\Curl\bu = \ba + \bc\times\bx
\eeq
where $\ba$ and $\bc$ are constant vectors. After substituting \eqref{curlu} into \eqref{curl1} and upon integration we obtain 
\beq\label{divu}
\Div\bu = \frac{2\mu}{2\mu+\lambda}\bc\cdot\bx + C
\eeq
where $C$ is a constant. Therefore, by using the obtained  information we can characterize the field $\bu(\bx)$ through its  Helmholtz decomposition  
\beq\label{Helm}
\bu=\nabla\varphi + \Curl\bomega
\eeq
where the scalar field $\varphi$ and the solenoidal vector field $\bomega$ 
can be found from the equations
\beq \label{divu11}
\left\{
\begin{array}{lll}
\displaystyle\D\varphi = \frac{2\mu}{2\mu+\lambda}\bc\cdot\bx + C\\
\displaystyle\D\bomega = - \ba - \bc\times\bx. 
\end{array}
\right. 
\eeq
We can therefore conclude that, by finding constants $(\ba,\bc,C)$ that are compatible with the boundary conditions, one can specify the compatible component of $\V_{\hspace{-1pt}\tau}(\bx,t)$ that is not affected by the relaxation mechanism implied in \eqref{div-curl-equi}. 

Although it cannot be excluded that this mechanism might leave some other compatible component of $\V_{\hspace{-1pt}\tau}(\bx,t)$ unrelaxed, the examples discussed in this work show that this does not occur. The only component of $\V_{\hspace{-1pt}\tau}(\bx,t)$ involved in the relaxation process is the one carrying incompatibility. Conversely, the compatible component, which remains unaffected by relaxation, is of the form $\sym(\Curl\U) = \sym\nabla\hat\bu$, with $\hat\bu$ exactly given by \eqref{divu11}.

\subsection{Reduced description \label{casestudy}}
 
Here we demonstrate how the surface deposition problem can be explicitly solved without any reference to the micro-displacement tensor $\V_{\hspace{-1pt}\tau}(\bx,t)$ in the physically relevant special case when $\gamma = 0$ and $\alpha = \infty$ simultaneously. Such reduced description is possible due to the complete decoupling of micro and macro problems which are, of course, tightly coupled in the general case.

To focus on the decoupling issue we simplify the problem even further by assuming that the evolution of the deposition surface is not determined self-consistently, but is instead prescribed independently. In other words, we assume that the function $\vartheta(\bx)$ is given. In this limited setting, the six components of the inelastic strain $\beps_p(\bx)$ are fully determined at the moment of deposition through six boundary conditions. Suppose that three of them impose  the  ``generalized forces''  $\bsigma_a$ 
\beq\label{extraconds}
\Omega_t: \qquad
\begin{array}{lll}
 \bsigma_{\tau}(\bx,\vartheta(\bx))  = \bsigma_a(\bx)
\end{array}
\eeq
while the other three 
\beq\label{extraconds1}
\Omega_t: \qquad
\begin{array}{lll}
\bu(\bx,\vartheta(\bx)) = \0
\end{array}
\eeq
imply that the displacements attributed to material points become meaningful only  after the instant of their deposition. Note that since in this oversimplified setting the inelastic strain $\beps_p(\bx)$ remains ``frozen''  after  the instant of deposition, we also have $\V_\tau=\V_\tau (\bx)$ which represents the decoupling of the  the micro-problem. The remaining macro-problem   reduces to finding the evolution of the displacement field $\bu(\bx,t)$.

Note next that the  field $\beps_p(\bx)$ can be eliminated by switching to an incremental formulation. Specifically, given that 
\beq
\bsigma(\bx,t)=\mathbb{C}\beps_e(\bx,t)
\eeq
and   
\beq
\beps_e(\bx,t)=\sym\nabla\bu(\bx,t)-\beps_p(\bx)
\eeq 
we can  write an incremental constitutive relation in the form
\beq
\dot\bsigma(\bx,t)=\mathbb{C}(\sym\nabla\dot\bu(\bx,t)).
\eeq
The associated incremental equilibrium equation is 
\beq\label{incrpb1}
\cB_t: \qquad
\Div\dot\bsigma(\bx,t) = \0. 
\eeq
To obtain the incremental boundary condition we differentiate the identity
\beq
\bsigma(\bx,\vartheta(\bx)) = \bbar\bsigma(\bx) 
\eeq 
with
\beq\bbar\bsigma(\bx) = \bsigma_a(\bx) + \bs_n \otimes \bn + \bn \otimes \bs_n - (\bs_n \cdot \bn)\bn \otimes \bn
\eeq
to obtain
\beq\label{incrpb11}
\Omega_t: \qquad
\dot\bsigma(\bx,\vartheta(\bx))\bn(\bx) = D(\bx)\Div\bbar\bsigma(\bx)
\eeq
where $D$ is the prescribed normal velocity of the deposition  surface. 

One can see that the ensuing  problem for the incremental displacement $\dot\bu(\bx,t)$ is purely elastic.  After solving a  sequence of such incremental problems,  one   can reconstruct  the time dependent stress and displacement fields at time $t$ using the relations:
\beq\label{increqns}
\left\{
\begin{array}{lll}
\displaystyle\bsigma(\bx,t) = \bbar\bsigma(\bx) + \int_{\vartheta(\bx)}^t\dot\bsigma(\bx,\tau)\,d\tau\\
\displaystyle\bu(\bx,t) = \int_{\vartheta(\bx)}^t\dot\bu(\bx,\tau)\,d\tau.
\end{array}
\right. 
\eeq
This completes the solution of the macro-problem, which is then indeed fully decoupled from the problem of identifying  the associated micro-displacement tensor.

To address the latter, we can use the  obtained function $\bu(\bx,t)$ and the  six closure conditions \eqref{extraconds} and \eqref{extraconds1}  to  
recover the distribution of the  ``frozen'' inelastic strain 
\beq\label{59}
\beps_p(\bx) = \mathbb{C}^{-1}\bbar\bsigma(\bx) - \sym(\nabla\bu(\bx,\vartheta(\bx))).
\eeq
Note that in view of \eqref{increqns} the obtained  function $\beps_p(\bx)$  depends non-locally on the controls $\bs_n(\bx)$ and $\bsigma_a(\bx)$. To show this more explicitly, we differentiate \eqref{extraconds}$_2$ to obtain 
\beq \label{591}
\nabla\bu(\bx,\vartheta(\bx)) = -\dot\bu(\bx,\vartheta(\bx))\otimes\nabla\vartheta(\bx).
\eeq
Equation \eqref{591} suggests that the inelastic strain at a point $\bx$ depends not only on the controls applied in this point at the instant of deposition, but also on the value of the incremental macro-displacement, which in turn depends on the entire prior deposition history. Accordingly, the  acquired incompatibility $\bet = \Curl\Curl\beps_p$ also inherits the non-local dependence on the applied controls. For instance, to ensure that  $\bet = \0$ and therefore the grown body is free of residual stresses, the protocol of applying the ``active''  stress $\bbar\bsigma$ must be specifically engineered.  
 
After the decoupled incremental  macro-problem is solved,  the associated micro-displacements described by the ``frozen''  tensor field $\Vtau(\bx)$ can be reconstructed by inverting the differential relation 
\beq \label{5912}
\beps_p (\bx)= -\sym(\Curl(\Vtau (\bx)\hodge \bn(\bx)))
\eeq
where the field $ \bn(\bx)$ is assumed to be  known. While we are not going to address this problem here in its full generality, our examples show how this can be done in special geometries.

\subsection{Coupled problem \label{casestudy1}}

We have seen that, in a special limiting case, the incremental macro-problem can be decoupled from the ``frozen'' micro-problem and solved in a fairly general setting. Such analytical transparency, however, is the exception rather than the rule. The fully coupled problem, incorporating for instance a self-consistent description of surface evolution, or addressing the mechanism of relaxation of the accumulated incompatibility, can be expected to pose significant analytical challenges. 

In particular the decoupling of the macroscopic description would not be usually possible, because the growth process is ultimately of micro-scale origin, involving active mechanical interactions which do not reduce to the action of  surface-associated active ``agents'', but also involve bulk evolution driven by the corresponding distributed volumetric ``active'' agents.  The presence of micro-scale mechanical controls/constraints leads to the emergence of  inelastic deformations and incompatibility encoded in the evolution of the micro-displacement tensor, which is no longer fully described by the deposition protocol itself. 

In the next sections we present a detailed study of several special problems where, due to the assumed radically simplified geometry, one can explore the effect of different active protocols and illustrate various conceptually challenging features of the proposed general theory.

\section{Activity assisted solidification\label{Sec-solid}}

As a first illustration of our approach,  consider a process of crystallization  from a liquid state interpreted  as a growth of a non-hydrostatically stressed elastic crystal from its melt. While this problem in its classical setting  is generally regarded as well understood, see for instance, 
\cite{Gibbs,
Sekerka-Cahn-2004, 
HobbsOrd, 
KaganovaRoitburd,
Lifshits1952, 
Levitas1998, 
Morris2017, 
Truskinovsky1984, 
Truskinovsky1983, 
frolov2010,
Green1980, 
Ghoussoub2001, 
Bowley1992, 
Myhill2025, 
Nye1967},  
our focus will be on the relatively new question regarding the potential role played by active ``agents'' on the development of incompatibility during crystallization. For analytical transparency, our analysis will be restricted to isotropic solids and will be carried under the assumptions of physically and geometrically linear elasticity. To emphasize the role of micro-mechanical actions in  the development of incompatibility, we also drastically simplify the thermal part of the problem by considering crystallization as an isothermal process, while neglecting in this way important aspects  of solidification process related to thermo-mechanical coupling \cite{YavariSafa2024}.

\subsection{Problem setting}

\begin{figure}[h!]
    \centering
    \includegraphics[width=0.55\linewidth]{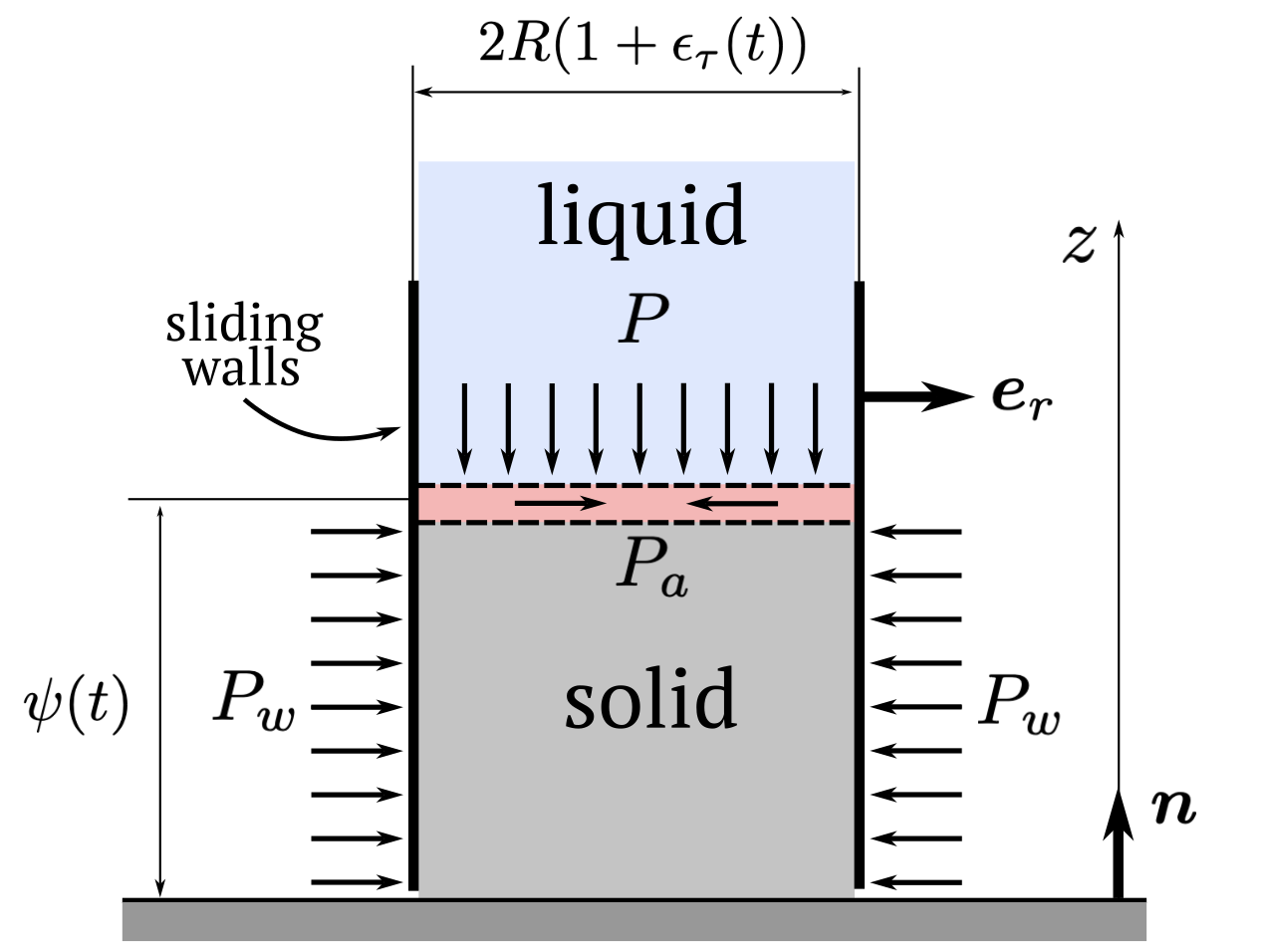}
    \caption{\label{Gibbs} Schematic representation of a solid under non-hydrostatic stress,  in contact with its melt. Vertical sliding walls   enforce homogeneous lateral deformation, with tractions $\bs_w=-P_w\be_r$. The   liquid  experiences a hydrostatic pressure $\bs_n = -P\bn$ with ($P>0$). The growth front, positioned at $\psi(t)$,   carries  controllable ``active" surface stress $\sigma_a=-P_a$.}
\end{figure}

Consider a solid body of cylindrical shape whose height $\psi(t)$ is a function of time due to the process of surface crystallization from a melt, see Fig. \ref{Gibbs}. Suppose that the material coordinates in the reference configuration of a solid body are 
\beq
\bx=r\be_r + z\bk
\eeq 
where $z\in(0,\psi(t))$ is the axial coordinate, $r\in(0,R)$ is the radial coordinate and $\be_r$ is the unit vector in the radial direction. Assume further that the layering of deposited material is horizontal with $\bn(\bx)\equiv\bk$ and that the position of the deposition surface at time $t$ is   $z=\psi(t)$.

We employ a semi-inverse approach and we introduce several assumptions constraining the nature of the emerging deformation. Specifically, we assume that the macro-displacement vector field is of the form  
\beq\label{ucylax}
\bu(\bx,t) = u(z,t)\bk + r\,\epsilon_{\tau}(t)\be_r(\bx). 
\eeq
Under this assumption, the macroscopic deformation of the growing cylindrical  solid body can be represented as an elongation/shortening in the axial   direction described by the $z-$dependent function $u(z,t)$,  together with a superimposed uniform shrinking - expansion in the tangential direction  described by a $z-$independent strain variable $\epsilon_{\tau}(t)$. In this way we also assume implicitly that at   $z = 0$, only the vertical displacements are constrained by the rigid foundation. 

Our assumptions regarding the structure of the micro-displacements can be formulated in the form of an ansatz
\beq
\Vtau(\bx,t)=v(z,t){{\P_{\bk}}}
\eeq
where ${\P_{\bk}}=\I-\bk\otimes\bk$ and  where the sub-continuum nature of the function $v(z,t)$ should be understood in the sense of \eqref{Vdiscrete}. In terms of \eqref{Vtau}, we are effectively assuming that $v_{11}=v_{22}=v(z,t)$ and $v_{12}=0$. The corresponding inelastic strain is then 
\beq\label{plast-cyl}
\beps_p(\bx,t)= - v'(z,t){\P_{\bk}}
\eeq
where we used the notation $'=\partial_z$. The corresponding elastic strain can be then written in the form 
\beq\label{epselacyl}
\beps_e(\bx,t) = (v'(z,t) + \epsilon_{\tau}(t)){\P_{\bk}} + u'(z,t)\bk\otimes\bk. 
\eeq 
Note next that due to the imposed kinematic constraint \eqref{ucylax}, the elastic stress tensor also has a special structure
\beq
\bsigma =\sigma_{\tau}(z,t){\P_{\bk}} + \sigma_{z}(z,t)\bk\otimes\bk
\eeq  
where $\sigma_{\tau}$ is in-plane component   and $\sigma_z$ is the normal, axial component. The off-diagonal  stresses are clearly absent  in our  simplified setting, while in-plane stresses have equal radial and hoop  components, $\sigma_r=\sigma_{\vartheta}=\sigma_{\tau}$. 
 
In what follows we neglect  body forces ($\bb=0$) and we assume that on the deposition/growth surface (with normal $\bn\equiv\bk$) tractions are hydrostatic
\beq
\bs=s_n\bn. 
\eeq
On the sliding walls of the cylindrical piston (with normal $\be_r$) tractions that are compatible with our other assumptions are necessarily radial, 
\beq
\bs=s_w\be_r. 
\eeq
We assume that the functions $s_n(\psi)$ and $s_w(\psi)$ are prescribed  and their magnitudes depend on the position of the growth surface. We can now write 
\beq
\sigma_{z}(\psi(t),t) =s_n(\psi(t))
\eeq
where 
 \beq
\sigma_{z} = (2\mu+\lambda)u' + 2\lambda (v'+\epsilon_{\tau}).
\eeq 
On the other hand, in view of  the constrained kinematics  \eqref{ucylax}, the loading in the transverse direction should be understood in the averaged (integral) form \cite{Palmov1967}. Specifically we have effectively assumed that  
\beq\label{Palmov1}
\int_0^{\psi(t)} \sigma_{\tau} (z,t)\,dz = \psi(t)s_w(\psi(t))
\eeq 
where
 \beq
 \sigma_{\tau} = \lambda u' + 2(\mu+\lambda)(v'+\epsilon_{\tau}). 
 \eeq 
We can then conclude that the function $s_w(\psi)$ represents the averaged  lateral tractions exerted on the body by the surrounding rigid loading device, see Fig.\ref{Gibbs}. For instance, in the case $s_w=0$,  Eq. \eqref{Palmov1} represents a continuous analogue of the condition \eqref{epsavdiscr}, which we used to describe the absence of lateral loading on a discrete stack of unevenly pre-stretched layers. 

To characterize our active ``agents'', we make the matching assumptions regarding the structure of the bulk and surface ``generalized forces'' featured in the expression of external power \eqref{power}.  Specifically, the corresponding tensor fields are assumed to be of the form: 
\beq
\textbf{B}_a(\bx,t)=B_a(z,t) {\P_{\bk}}, \qquad \bsigma_a(\bx)=\sigma_a(z,t) {\P_{\bk}}
\eeq
where $B_a(z,t)$ and $\sigma_a(z,t)$ are externally prescribed scalar fields representing microscopically applied sub-continuum controls.

\subsection{Evolution equations}

Our analysis will be  performed   under the assumption  that mechanical equilibrium holds for macro-displacements  in the sense that \eqref{Pbulk1},\eqref{Pbnd1} hold throughout the whole process of crystal growth.  We  assume further  that $\alpha>0$ in \eqref{Pbulk2} and $\beta>0$ in \eqref{eqD3}, while  we set $\gamma=0$ in \eqref{Pbnd2}. 

In view of the assumptions made above, the mechanical macro-equilibrium in the axial direction for $\sigma_{z}(z,t)$ takes the form  
\beq
\sigma_{z}' =  0 \qquad z\in(0,\psi(t))
\eeq
while the only macro-equilibrium equation involving $\sigma_{\tau}(z,t)$ is \eqref{Palmov1}. The evolution equation for $v(z,t)$ in the bulk reduces to 
\beq \label{B1}
\sigma_{\tau}' + B_a = \alpha\dot v \qquad z\in(0,\psi(t)). 
\eeq
Note that  the dynamics described by \eqref{B1} terminates at micro-equilibrium where  $\sigma_{\tau}' = - B_a$. This suggests  an interpretation of $B_a$ as  a force  of dry friction type preventing  the infinitesimal neighboring layers from relative horizontal  sliding.  For instance, if the tangential stress $\sigma_{\tau}$ is discontinuous across a   planar surface $z=a$  so that $\sigma_{\tau}(z)=\sigma_{\tau}^-+(\sigma_{\tau}^+-\sigma_{\tau}^-)H(z-a)$, where $H(z)$ is the Heaviside function and $\sigma_{\tau}^{\pm}$ are given constants, the implied  micro-equilibrium   is reached for $B_a=(\sigma_{\tau}^--\sigma_{\tau}^+)\delta(z-a)$ where $\delta$ is the Dirac delta. The corresponding ``generalized force'' $\B_a=B_a{\P_{\bk}}$ is then represented by a singular tensor field of the type discussed in   \cite{Angelillo, Silhavy,ppg205,PandeyGupta21}.
 
To simplify further the treatment, we now assume that $\alpha=\infty$ which means that $v(z,t)=v(z)$. The corresponding micro-displacement field is then effectively ``frozen'' at the instant of deposition. Such ``freezing'' must be, of course,  supported by a distribution of the generalized forces $B_a(z)$  which would then  provide  an effective  ``shear glue'' needed  to  maintain the acquired level incompatibility. 

Given our assumptions, the dynamic equation governing the evolution of the  function   $\psi(t)$  takes the form 
\beq\label{Dpsi}
W_a - \bbar{W}  + s_n \bbar{u'} + 2 \sigma_a \bbar{v'} = \beta \dot{\psi}(t)
\eeq
where we  continue to  use the notation 
\beq
\bbar A(\bx)=A(\bx,\vartheta(\bx)).
\eeq  
We re call that in \eqref{Dpsi}, the function $s_n(\psi)$ prescribes a (passively) controlled axial traction. Instead, the control function  $\sigma_a (\psi)$ can be viewed as characterizing the active ``agents'' at the phase boundary, see Fig.\ref{Gibbs}. The function $W_a(\psi)$, formally defined  in our general theory as the energy of the incoming material, can be interpreted in the present case of crystallization from a melt as the energy density of the liquid phase at the instant of deposition. Since the liquid phase is loaded hydrostatically by the pressure $P$, and therefore $s_n=-P$, the simplest assumption would be that the arriving liquid has the same energy density as  the corresponding solid loaded hydrostatically by the same pressure $P$.  Under this assumption  
\beq \label{lambda1}
W_a = \frac{3(1-2\nu)}{2 E}P^2
\eeq
where we have introduced  the Young modulus $E$ and the Poisson ratio $\nu$, related to Lamé constants through
\beq \label{lambda}
\mu = \frac{E}{2(1+\nu)}, \qquad
\lambda = \frac{E\nu}{(1+\nu)(1-2\nu)}. 
\eeq
The function $ W $ in \eqref{Dpsi}, describing the elasticity of the solid phase, has a standard form  
\beq
W = \mu(2 e_{\tau}^2+e_z^2) + \frac{\lambda}{2}(2 e_{\tau}+e_z)^2
\eeq
where for the tangential and axial components of the elastic strain, we substitute $e_{\tau} = v' + \epsilon_{\tau}$ and $e_z = u'$, respectively. The chosen representation of the energy of the solidifying liquid in \eqref{lambda1} is, of course, highly schematic, as it completely neglects the existence of an energy well corresponding to the liquid phase (see, for instance, \cite{GraTru23}). Nevertheless, as we show below, even such a simplified formulation is sufficient to illustrate the influence of the micro-mechanical controls encoded in the function $\sigma_a(\psi)$.

\subsection{Solution of the dynamic equations} 

We begin by writing   the full system  of dynamic equations  for  the  four unknowns functions $u(z,t),v(z),\psi(t),\epsilon_{\tau}(t)$.  
It includes
the equilibrium equations in the bulk: 
\beq\label{79}
z\in (0,\psi(t)): \quad
\left\{
\begin{array}{lll}
\sigma_{z}' =  0\\
\int_0^{\psi(t)}\sigma_{\tau}(z,t)\,dz = \psi(t)s_w(\psi(t))
\end{array}
\right. 
\eeq
supplemented by  the constitutive equations in the bulk 
\beq\label{78}
z\in (0,\psi(t)): \quad
\left\{
\begin{array}{lll}
\sigma_{\tau} = \lambda u' + 2(\mu+\lambda)(v'+\epsilon_{\tau})\\
\sigma_{z} = (2\mu+\lambda)u' + 2\lambda (v'+\epsilon_{\tau})
\end{array}
\right. 
\eeq
together with the boundary-initial conditions conditions
\beq\label{planar}
\left\{
\begin{array}{lc}
\psi(0) = \psi_0 \\
u(0,t)=0 \\
v(0,t)=v_0 \\
\bbar\sigma_{z}(\psi(t),t) = s_n(\psi(t)) \\
\bbar\bbsigma(\psi(t),t)= \sigma_a(\psi(t)). 
\end{array}
\right. 
\eeq
Here $\psi_0$ is the thickness of an initial basal layer which is instantaneously created under zero passive and active loading; $v_0$ is the initial value of the micro-displacement parameter at deposition.
Since we deal with a free boundary problem, we must include  the evolution equation for the deposition/growth  surface: 
\beq\label{80}
z=\psi(t): \quad
W_a - \bbar{W}  + s_n \bbar{u'} + 2 \sigma_a \bbar{v'} = \beta \dot{\psi}(t). 
\eeq
We assume that the   parameters $\psi_0$, $v_0$, ${W}_a(P)$ and the control functions  $s_n(\psi)$, and $s_w(\psi)$ are all prescribed, however, as we show below,  they  cannot be chosen completely arbitrarily.

Substituting  \eqref{78} and \eqref{planar}$_{4,5}$ into \eqref{79} we  immediately obtain an  explicit representation of the ``frozen'' micro-displacements characterized by the function $v(z)$
\beq\label{dv}
v'(z) = - \frac{1-\nu}{E}\left(s_w(z)-\sigma_a(z)+\cF(z)\right)
\eeq
where we have introduced the auxiliary function
\beq\label{FF}
\cF(z) = \int_0^z\frac{s_w(\tilde z)-\sigma_a(\tilde z)}{\tilde z}\,d\tilde z. 
\eeq
To avoid the formation of a singularity at $z = 0$, the function $\cF(z)$  must remain bounded as $z \to 0$; physically, this means that the controls should be applied sufficiently gradually. It is also clear from \eqref{FF} that in the special case when the passive axial forces and the active lateral forces on the deposition surface are equal  
\beq\label{sigmacomp}
\sigma_a(\psi) = s_w(\psi)
\eeq
the micro-displacements are absent and, in particular, no incompatibility is generated during the process. Moreover, one can see that $v' = 0$ if and only if $s_w(z) - \sigma_a(z) = c/z$ for some constant $c$, so to avoid a stress singularity at the origin we must necessarily have $c = 0$. Consequently, a microscopically engineered active deposition protocol satisfying \eqref{sigmacomp} can be termed \textit{compatible}, as it ensures that any solid body grown under these conditions is stress-free when passive tractions are removed.

Given our simplifying assumptions, we can also find explicit expressions for the transversal strain 
\beq\label{epstau}
\epsilon_{\tau}(t) = \frac{(1-\nu)}{E}\left(\cF(\psi(t))+s_w(\psi(t)) - \frac{\nu}{(1-\nu)} s_n(\psi(t))\right)
\eeq
and for the axial strain 
\beq\label{du}
u'(z,t) = \frac{1}{E(\nu-1)}\left((1+\nu)(2\nu-1)s_n(\psi(t)) + 2 E \nu(v'(z)+\epsilon_{\tau}(t))\right)
\eeq
where it is implied that  $z \leq \psi(t)$ while the function $\psi(t)$   is determined from  \eqref{Dpsi}.

\subsection{Regime diagram} 

Let us now consider a special case in which the role of different deposition strategies can be assessed quantitatively. To begin with, in order to avoid the formation of a singularity at $z = 0$, we divide the deposition process into two stages: a ``nucleation'' stage, during which an initial layer of thickness $\psi_0$ is instantaneously created under zero loading, and therefore
\beq\label{loading0}
\left\{
\begin{array}{lll}
s(z) = 0\\
\sigma_a(z) = 0\\
s_w(z) =0
\end{array}
\right. 
\qquad (0<z<\psi_0)
\eeq
and a  ``growth'' stage  when the crystalization advances  under special loading conditions (see Fig.\ref{Gibbs})
\beq\label{loading1}
\left\{
\begin{array}{lll}
s(z) = - P\\
\sigma_a(z) = - P_a\\
s_w(z) = - P_w
\end{array}
\right. 
\qquad (\psi_0<z). 
\eeq
In what follows, it will be convenient to use  dimensionless parameters   
\beq
 p_a  = P_a/P, \qquad p_w  = P_w/P 
\eeq
with the goal of  constructing a regime diagram in the $(p_a, p_w)$ space. Within this setting, we have  $\cF(z)=0$ during the   ``nucleation'' stage, and 
\beq\label{cF0}
\cF(z)={P}({p_a}-p_w)\log(z/\psi_0)
\eeq
during  the  ``growth'' stage. The evolution  of the phase boundary $\psi(t)$ is determined by solving   the   problem
 \beq \label{evolutionpb}
\left\{
\begin{array}{lll}
\beta \dot\psi(t) = \cG(\psi(t))\\
\psi(0) = \psi_0
\end{array}
\right. 
\eeq
where
\beq
\cG(\psi) = 2{p_a} (1-\nu)\frac{{P}^2}{E}\left(A + ({p_a}-p_w)\log\left(\frac{\psi}{\psi_0}\right)\right). 
\eeq
In \eqref{evolutionpb} we  also introduced  the notations
\beq
A = \frac{S(\nu)}{2{p_a}} + \frac{{p_a}-2p_w}{2}, \qquad S(\nu)=\frac{2-3\nu}{1-\nu}. 
\eeq
Note first that the ODE  \eqref{evolutionpb} has a single fixed point
\beq\label{psievo}
\psi_e = \psi_0 e^{A/(p_w-{p_a})}
\eeq
where $\cG(\psi_e)=0$.  It is stable if  
\beq
\cG'(\psi_e) = \frac{2 {P}^2}{E\psi_e}{p_a}(1-\nu)({p_a}-p_w)<0 
\eeq
or, equivalently, if
  \beq
p_w > {p_a}.
\eeq
We recall from \eqref{sigmacomp} that  the limiting case   $p_w = {p_a}$ corresponds to  \textit{compatible}  growth. 

\begin{figure}[h!]
    \centering
    \includegraphics[width=1\linewidth]{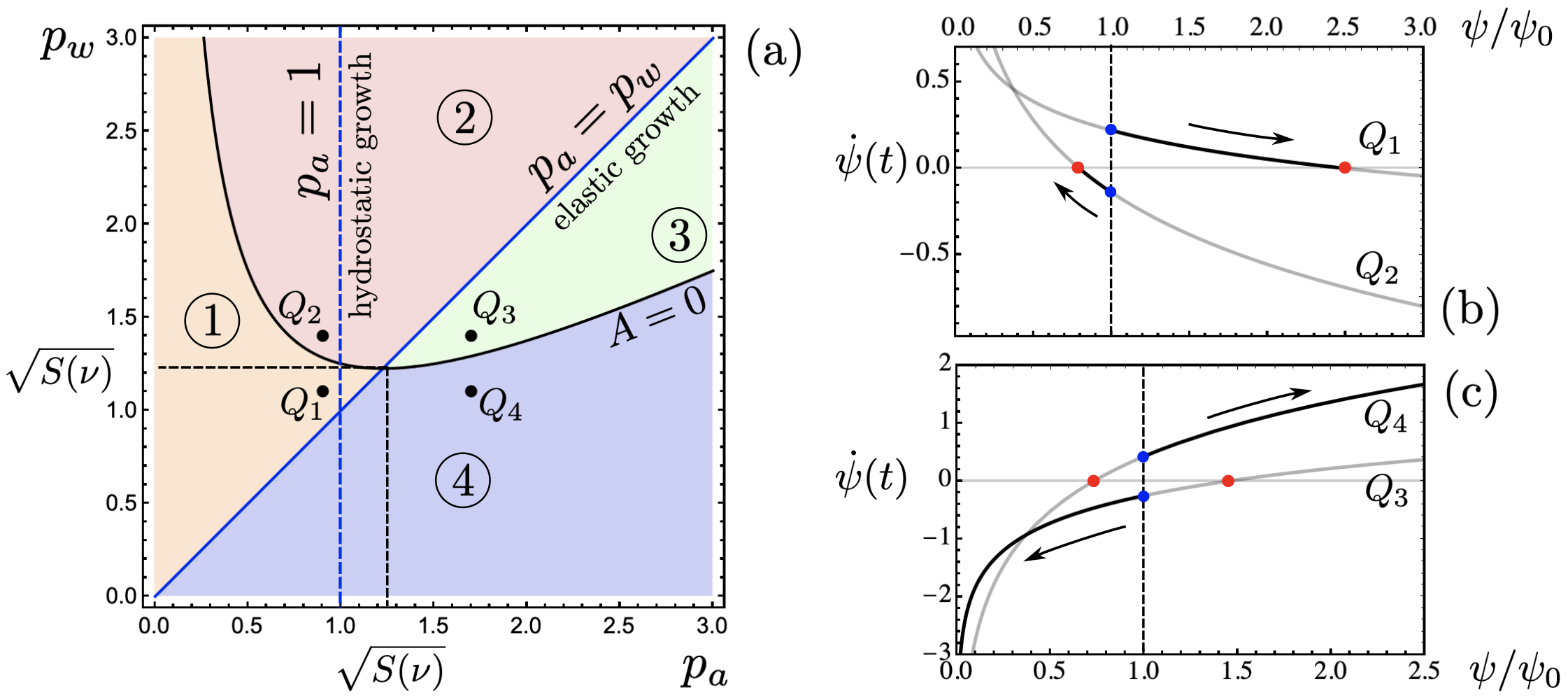}
    \caption{\label{papw} (a) Regime diagram in the space  of parameters $({p_a}, p_w)$ for the   evolution problem $\dot\psi = \mathcal{G}(\psi)$ with $\psi(0) = \psi_0$, see   \eqref{evolutionpb}. In regions 1 and 2,   $\cG'(\psi_e) < 0$, producing  asymptotically stable regimes. In   regions 3 and 4, $\cG'(\psi_e) > 0$, indicating unstable regimes.
(b)–(c) The trajectories of interest on the phase  plane $(\dot\psi, \psi / \psi_0)$. Here    $\psi_0$ is the initial location of the growth surface (blue dot).
(b) Stable trajectories (regions 1 and 2) converge asymptotically to equilibria (red dots), which  are either above ($Q_1$) or below ($Q_2$) the initial state.
(c) Unstable trajectories (regions 3 and 4) either extending to infinity ($Q_4$) or collapsing to zero ($Q_3$). All plots are obtained with Poisson ratio $\nu = 0.3$. The choices  of the parameters ${P},E,{p_a},p_w$ only contribute to  rescaling  of the trajectories shown in (b) and (c).}
\end{figure}

We are now in a position to identify distinct dynamic regimes in the parameter plane $({p_a}, p_w)$ (see Fig.~\ref{papw}).  Specifically, we observe  stable growth  ($\psi_0<\psi_e<+\infty$) in region 1, where $p_w>{p_a}$ and  $A>0$, and therefore in terms of the parameters: 
\beq
\frac{1}{2}\left( \frac{S(\nu)}{{p_a}} +{p_a}\right) > p_w > {p_a}. 
\eeq
In region 2, where  $p_w>{p_a}$ and  $A<0$  so that 
\beq
p_w>\text{max}\left(\frac{1}{2}\left(\frac{S(\nu)}{{p_a}} +{p_a}\right), {p_a}\right)
\eeq
we observe  stable resorption ($0<\psi_e<\psi_0$).  Regions 3 (where $A > 0$ and ${p_a} > p_w$) and 4 (where $A < 0$ and ${p_a} > p_w$) correspond to unstable regimes, because  in both cases $\cG'(\psi_e) > 0$.

We now discuss  in more detail   two special  regimes, emphasizing the role played by the microscopically generated control, represented by the active tangential stress $\sigma_a=-p_a P$. 

\medskip

$\bullet$ \textbf{Compatible (elastic) growth.} This is the classically studied regime  with  no prestress   accumulated in the body, see the line ${p_a}=p_w$ in Fig.\ref{papw}(a).  In this case the fixed point \eqref{psievo} becomes degenerate, as it is displaced  either  to infinity or to  zero depending on whether $A>0$ or $A<0$, respectively. Specifically, when 
\beq
{p_a}=p_w<\sqrt{S(\nu)}
\eeq
the function $\psi(t)$ blows-up with time. Instead, it collapses to zero when 
\beq
{p_a}=p_w>\sqrt{S(\nu)}.
\eeq
The critical threshold ${p_a}=p_w= \sqrt{S(\nu)}$ will then define a  boundary between unbound growth and full resorption. One can see that the presence of the  microscopic  mechanical activity can, in principle,  reverse the direction of a  phase transformation from solidification to melting. 

Note that while the  \textit{compatible} phase transformation paradigm, already employed by Gibbs in his classical study of solid–liquid interface stability \cite{Gibbs}, is an analytically transparent choice, it is likely a useful abstraction rather than a physically realizable mechanism. Indeed, to avoid the emergence of incompatibility under the assumption of \textit{compatible} growth, the hydrostatically stressed liquid elements must be transformed into non-hydrostatically stressed solid elements by some physically obscure  active ``agents'' executing the required pre-stretch. In this perspective  the development of at least some incompatibility appears physically more natural, in particular, given the  random orientation of the crystallites during the  solidification from a melt. We also observe that {\it compatible} growth does not allow the system to attain a configuration of finite size: this limitation becomes a crucial obstacle when applying the model to biological systems, where, interestingly, the connection between the emergence of finite-sized configurations and the inherently incompatible nature of growth has been recognized (see, e.g., \cite{AEGZ24, AEGZ25}).

\begin{figure}[h!]
    \centering
    \includegraphics[width=1\linewidth]{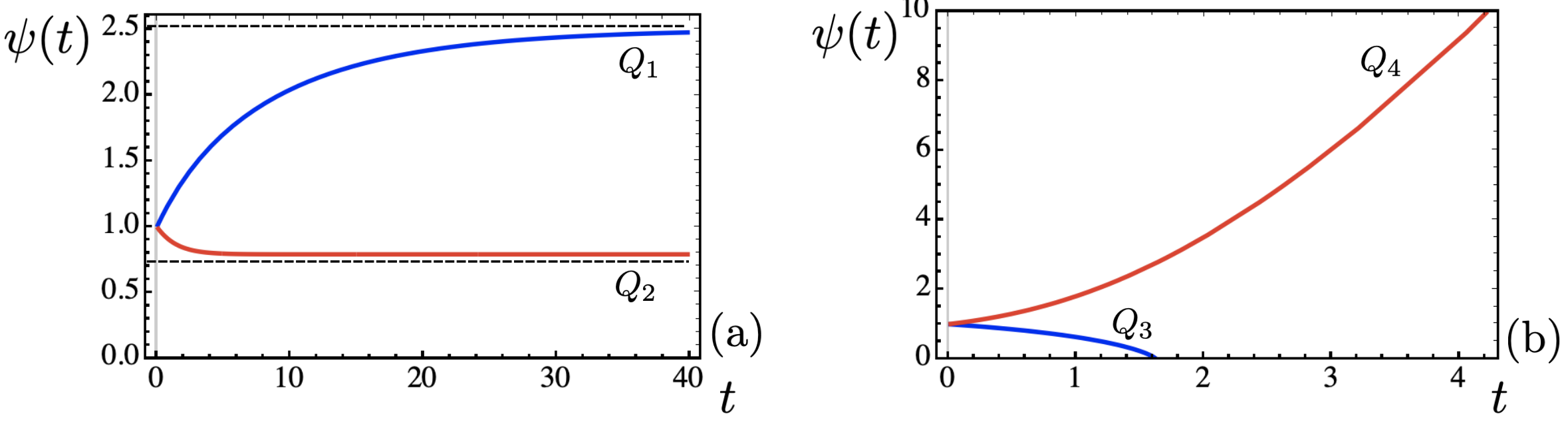}
    \caption{\label{evo} (a) Evolution of growing surfaces $\psi(t)$ represented by the solutions  of the problem  $\dot\psi = \mathcal{G}(\psi)$ with initial condition $\psi(0) = 1$, and parameters corresponding to the points $Q_1 = (0.9, 1.1)$ and $Q_2 = (0.9, 1.4)$ in Fig.~\eqref{papw}. In both cases, $\psi(t)$ converges to a finite equilibrium, which is higher than the initial value in the first case  (solidification) and lower in the second (melting).
(b) Behavior  of the function $\psi(t)$ corresponding to  points $Q_3 = (1.7, 1.4)$ and $Q_4 = (1.7, 1.1)$ in Fig.~\eqref{papw}. In these cases, no stabilization is possible: $\psi(t)$ collapses to zero in the former case, while it blows up in the latter case. Parameters are fixed at $\nu = 0.3$ and ${P} = E = 1$.}
\end{figure}
 
\medskip

$\bullet$ \textbf{Hydrostatic growth.} Another important regime is characterized by the condition $\sigma_a= s_n =-{P}$, represented  by the line ${p_a} = 1$ in Fig.~\ref{papw}, and describing a situation where the arriving solid elements are loaded  exactly as their parent fluid elements. In this case the  active micro stresses match their passive macro-analogs, ensuring that the deposited elements are loaded hydrostatically. One can say that the liquid retains the memory of its stress state as it transforms into the solid. While such an assumption seems natural \cite{KingFletcher,Trincher}, it clearly leads to the  development of incompatibility and accumulation of residual stresses.

By implementing the hydrostatic growth protocol, different dynamic regimes can be reached depending on the value of the parameter $p_w$, describing the lateral pressure on the walls of the cylinder. When $p_w < 1$ (lateral pressure is  lower  than the liquid pressure), the solid phase grows indefinitely. When $p_w \in (1, 2 + {0.5}/{(\nu - 1)})$, the system evolves toward a stable equilibrium size which exceeds the initial one (solidification). When  instead $p_w > [2 + {0.5}/{(\nu - 1)}]$, the  stabilization produces  system shortening (melting). 
These observations explain, for instance,  the peculiar behavior observed during ice melting,  as reported in \cite{Dash}. 

 The obtained results  are  also relevant for the description of  multicellular tumor spheroids \cite{Montel}, where lateral confinement,  known as  “stress clamping” (quantified by parameter analogous to $p_w$), was shown to slow down growth and even trigger resorption. Similar  mechanically mediated regulation of growth has also been documented for  healthy tissues \cite{Taber2009,Shraiman2005}.

\subsection{The final configuration}

The presence of accumulated incompatibility can be detected by residual stresses in the final configuration. In  the absence of applied macroscopic tractions, the equilibrium equations can be solved analytically, revealing that the residual stresses in a cylinder of the final height $\psi_f$ are of the form 
\beq\label{res-cyl}
\sigma^0_z = 0, \qquad\sigma_{\tau}^0(z) = \frac{E}{1 - \nu} \left( \langle \gamma \rangle - \gamma(z) \right)
\eeq
where 
\beq\label{res-cyl1}
\gamma(z)=-v'(z)
\eeq
is the tangential component of the inelastic  strain  \eqref{plast-cyl} while
\beq
\langle \gamma \rangle = \frac{1}{\psi_f} \int_0^{\psi_f}\gamma(z) dz
\eeq
stands for  the average accumulated inelastic  strain. Note an analogy  between   \eqref{res-cyl} and the corresponding result in our discrete theory  \eqref{epsavdiscr1}. 
 
The representation \eqref{res-cyl} suggests that, in the kinematically constrained case under consideration, a natural measure of incompatibility is
\beq\label{eta-cyl}
\eta(z) = \langle \gamma \rangle - \gamma(z) = v'(z) - \frac{v(\psi_f) - v(0)}{\psi_f}.
\eeq
Noting from \eqref{eta-cyl} that linear distributions of micro-displacements $v(z)$ are always compatible in this sense, we introduce the  ``compatible component"  of the micro-displacement field 
\beq
v_c(z) = v(0) + \frac{z}{\psi_f}\big(v(\psi_f) - v(0)\big)
\eeq
which allows us to isolate the ``incompatibility-related” component
\beq\label{vinc-cyl}
v_i(z) = v(z) - v_c(z)
\eeq
which is characterized by the property
\beq
\eta(z) = v_i'(z).
\eeq
Furthermore, because by definition $v_i(0)=v_i(\psi_f)=0$, it is clear  that $\eta(z) = 0$ if and only if $v_i(z)=0$. 

Observe next  that using these new definitions, we can rewrite the expression for the final elastic strain in the grown solid \eqref{epselacyl} in the form
\beq\label{epselacyl-2}
\beps_e (\bx) = \nabla\hat\bu (\bx) + v_i'(z){\P_{\bk}}, \qquad \hat\bu(\bx)  = u(z)\bk + r \hat\epsilon_{\tau}\be_r
\eeq
where we have set 

\beq
\hat\epsilon_{\tau} = \epsilon_{\tau}- \langle\gamma\rangle=  (P /E) \left(\nu-{p_w}(1-\nu)\right).
\eeq
The message behind such a rewriting   is that the uniform lateral micro-dilatation can be  absorbed into  macro-deformation, which is similar to what happens in the discrete case already,  see for instance Eq. \eqref{epsavdiscr}. 
 
To compute the distribution of incompatibility $\eta(z)$ when the body reaches the final size with $\psi(T_f)=\psi_f > \psi_0$, we first introduce useful notations.  Suppose that   the function $\gamma(z)$  is defined  by  \eqref{res-cyl1}. We use  the superscript ``$-$'' for the  ``nucleation'' layer at  $z \in (0, \psi_0)$ where $\gamma(z) = \gamma^-(z) = 0$. Accordingly, we identify by the superscript ``$+$''  the  ``grown'' layer at  $z \in (\psi_0, \psi_f)$ where $\gamma(z) = \gamma^+(z) \neq 0$. 

\begin{figure}[h!]
    \centering
    \includegraphics[width=\linewidth]{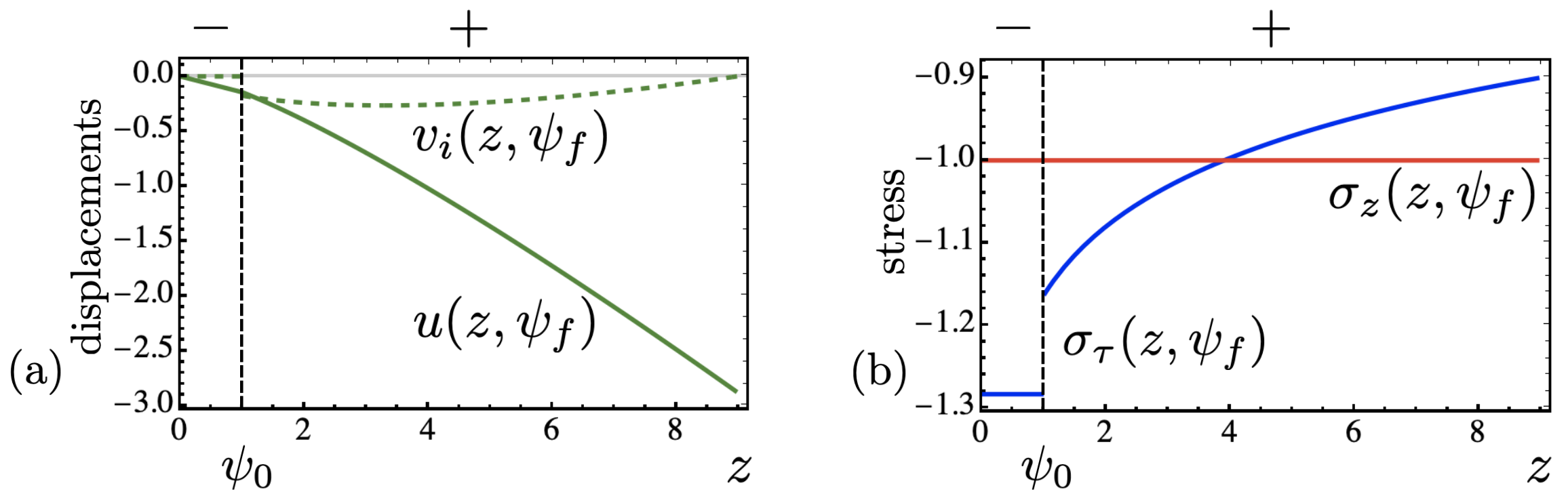}
    \caption{\label{dispstress}    (a) Macro displacement field $(solid line)$ and incompatible-micro displacement field $(dashed line)$;  (b) stress components in the axial direction $(\sigma_z)$ and in the horizontal direction $(\sigma_{\tau})$, in the basal layer ``-'' $(0,\psi_0)$  and in the grown region ``+'' $(\psi_0,\psi_f)$. Here $\psi_0=1$, $\nu=0.3$, ${P}=E=1$, and we have chosen  $({p_a},{p_w})$ corresponding to point $Q_1$ in Fig.\ref{papw}.}
\end{figure}

Note next  that  from  $v^-(0)=0$ we  obtain $v^-(z)=0$ in $(0, \psi_0)$, and by integrating \eqref{dv} with initial condition $v^+(\psi_0)=0$, we obtain $v^+(z)$ in $(\psi_0,\psi_f)$.  More specifically, we obtain  
\beq
v_i^-(z)=0,\qquad
v_i^+(z) = \frac{{P} (p_w-{p_a})(1-\nu)}{E} z\log\left(\frac{z}{\psi_f}\right).
\eeq
The corresponding  incompatibility takes the form
\beq \label{Ink3}
\eta^-(z)=0,\qquad
\eta^+(z) = \frac{{P} (p_w-{p_a})(1-\nu)}{E}\left(1+\log\left(\frac{z}{\psi_f}\right)\right). 
\eeq
To obtain the distribution of macro-displacements we  need to integrate \eqref{du} in $(0,\psi_0)$ with the boundary condition $u^-(0,T_f)=0$, giving 
\beq
u^-(z,T_f) ={P} z\frac{2{p_w}\nu -1 + 2({p_w} - {p_a})\nu\log(\psi_f/\psi_0)}{E}.
\eeq
and also \eqref{du} in $(\psi_0,\psi_f)$ with the boundary condition  $u^-(\psi_0,T_f) = u^+(\psi_0,T_f)$, giving 
\beq
u^+(z,T_f) = {P} z\frac{2{p_w} \nu - 1 + 2\nu({p_a}-{p_w})\log(z/\psi_f)}{E}. 
\eeq
The  functions $u^{\pm}(z,T)$ and $v_i^{\pm}(z)$ are  illustrated in Fig.\ref{dispstress}$(b)$.  

It is also instructive to reconstruct the  final   stress configuration. Thus, in the ``nucleation'' zone  
\beq
z\in(0,\psi_0):\qquad
\left\{
\begin{array}{lll}
\sigma_z^-(z,T_f) = -{P}\\
\sigma_{\tau}^-(z,T_f) = -{P}\left({p_w}+({p_a}-{p_w})\log\left(\psi_0/{\psi_f}\right)\right)
\end{array}
\right. 
\eeq
while in the ``grown'' zone  
\beq
z\in(\psi_0,\psi_f):\qquad
\left\{
\begin{array}{lll}
\sigma_z^+(z,T_f) = -{P}\\
\sigma_{\tau}^+(z,T_f) = -{P}\left({p_a}+({p_a}-{p_w})\log\left({z}/{\psi_f}\right)\right). 
\end{array}
\right. 
\eeq
Both distributions are illustrated in  Fig.~\ref{dispstress}(a). Observe that the axial stress $\sigma_z$ is uniform, which is consistent with the statically determined nature of the problem along the axial direction. In contrast, the tangential (horizontal) stress $\sigma_{\tau}$ is   inhomogeneous as long as ${p_a} \neq {p_w}$. Note also   a tangential stress discontinuity across the internal interface at $z = \psi_0$. When ${p_a} = {p_w}$, the stress distribution becomes homogeneous throughout the entire body even when  the configuration is non-hydrostatic,  which is the  behavior anticipated by Gibbs  \cite{Gibbs} in the absence of incompatibility.

\subsection{Discussion}

Our analysis  of a non-hydrostatically loaded  solid in contact with its melt presented in this section  can be viewed as an extension of Gibbs seminal study \cite{Gibbs}.  Here we discuss  the new elements contributed by our approach to the established picture of this phenomenon. 

While chemical aspects such as compositional differences between the coexisting solid and liquid phases, and the associated heat release, have been disregarded, we have proposed a more nuanced representation of the solid phase compared to previous studies. In addition to the elastic strain, we now also account for inelastic contributions; furthermore, we introduce a mechanism that regulates the strain incompatibility arising during solidification, by incorporating a description of the microscopic activity responsible for imposing pre-stress on the newly formed solid elements.

We recall that in Gibbs' classical treatment \cite{Gibbs}, the growth of the solid phase from its melt was effectively assumed to be compatible, or equivalently, purely elastic (see also \cite{HobbsOrd, Sekerka-Cahn-2004, GraTru23}). In contrast, our study focuses on the emergence of incompatibility enabled by an extended set of controls, involving the interplay among three quantities: the classical traction exerted by the fluid phase, $s = -P$ (the macroscopic axial stress); the lateral pressure, $s_w = -p_w P$ (the macroscopic lateral confinement stress); and the active surface (tangential) stress, $\sigma_a = -p_a P$. In these terms, Gibbs implicitly assumed that $p_a = p_w$, or equivalently, $\sigma_a = s_w$.

Our approach reveals a much broader repertoire of solidification scenarios compared to the conventional picture limited to compatible growth, which is represented by a marginally stable dividing line in Fig.~\ref{papw}. Under our expanded set of controls, the entire ${p_a}$ -- ${p_w}$ plane becomes accessible, allowing for a range of growth pathways and stability regimes which may be of particular relevance for living systems that are known to experience biologically functional liquid-solid transitions. In particular, microscale mechanical activity represented by the deposition pressure $p_a$ was shown to be a crucial player, able by itself to both initiate and inhibit growth. It was also shown that bounded and unbounded growth occur depending on whether the parameters $({p_a},p_w)$ fall into the corresponding regions in the phase diagram shown in Fig.\ref{papw}.  

In this respect, we recall that in his study, Gibbs observed that in the typical case of compatible growth with ${p_w} \neq 1$ (i.e., $s_w \neq -{P}$), even thermodynamically equilibrated solid-liquid interface is morphologically unstable. More generally, he concluded that a non-hydrostatically stressed solid in equilibrium with a hydrostatically stressed fluid is inherently unstable: his argument was that a stressed element of the solid surface can always dissolve into the fluid/melt and recrystallize elsewhere in hydrostatic conditions, thereby lowering the total energy. This instability mechanism was later examined more systematically in \cite{AsaroTiller} in a linearized setting, and in \cite{Grinfeld} within a geometrically and physically nonlinear framework.   More recently, it was shown that the same instability can also be detected by using  a novel equilibrium jump condition discovered in \cite{GrabovskyTruskinovskyARMA}. Translated into our language, this condition states that under the assumption of elastic growth with ${p_a}= {p_w}$, stable coexistence of a solid with a fluid is only possible when ${p_w} = 1$, that is, when the solid is hydrostatically stressed. Here we broaden the stability picture developed by Gibbs and his successors by allowing for inelastic growth, and we show that, in this case, stability may in principle be recovered even when ${p_w} \neq 1$, due to the presence of active ``agents'' able to impose a non-classical micro-stress $\sigma_{a}$ at the deposition surface.

We also mention that while the final size $\psi_e$ reachable in the solidification  process  is  ultimately selected dynamically, our analysis suggests that it  depends   only on the initial size of the body $\psi_0$. In other words, in contrast to models where growth leads to an intrinsic size selection, see  for instance \cite{AEGZ24,AEGZ25}), here due to the absence of an internal length scale, we observe  only a  finite size effect. An important observation, however,  is that the ultimate   stabilization takes place only  when $p_w \ne {p_a}$, that is, when growth is incompatible. Instead,   as we have seen,   along the compatibility line $p_w = {p_a}$, no such final stabilization occurs as the system either explodes  or collapses. 

To summarize, despite the simplifying assumptions made in our study, in particular the fact that we focus exclusively on mechanical aspects, we have provided a compelling evidence   that  the formation of  incompatibility is an essential part of solidification, with inelastic strains playing a fundamental role on the stability of the solid-liquid front. A natural conjecture emerging from our study    is that the realistic self-driven   solidification process can  be hardly ``elastic'' and therefore  the emergence of inelastic strains and the formation of incompatibility in such process is  unavoidable. The effective nature of 
active controls in a constrained cooling process remains to be understood through detailed microscopic studies, see for instance  \cite{Lutsko2006, Mullin2001,Yukalov1985,Svishchev1994, Yau2000}. A comprehensive  mechanical study of the solidification process in realistic technological systems accounting for the formation of residual stresses is also an important open problem, despite many past and ongoing efforts \cite{Perez2024, ASMHandbook,Chalmers1964,Hu1996,Kurz2023}.

\section{Winding with  pre-stretch\label{section-winding}}

Our second example concerns the technologically important problem of roll winding, where residual stresses  arise rather naturally due to   variable tension applied to  a wire, a  tape or a film \cite{Southwell}.   The challenge is to link the emerging residual stresses to particular winding protocol \cite{Zabaras95}.  

Recently, a problem of controlling stress distribution in wound rolls and other similar ``grown'' structures has received significant attention in the technical  literature aimed at the optimization of winding machines \cite{good2008winding,arola2007two, park2018effects,dehoog2007inverse,krysiak2021theoretical,Andreotti25}. In  most of these studies  the analysis was performed at what we would interpret as a micro-scale, in particular   the  thickness  of the attached film, tape or wire was explicitly resolved and  the emerging incompatibility  was represented by the relative slip of the attached layers. 

Our coarse grained approach offers a possibility to describe  the same   winding processes directly at the macro-scale, without the layering thickness explicitly resolved. In such approach, winding is interpreted as a surface deposition process, with the variable tension on the deposited segment of  tape, wire or film  being responsible for the emerging incompatibility.

\subsection{Problem setting}
  
\begin{figure}[h!] 
    \centering
    \includegraphics[width=0.7\linewidth]{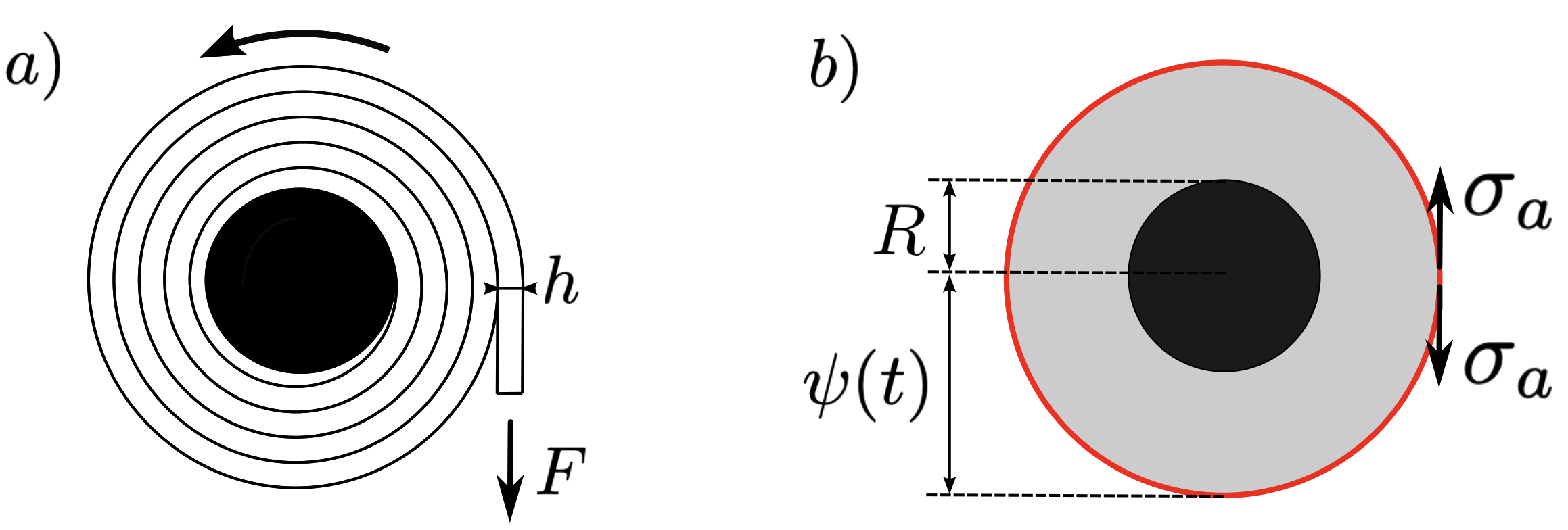}
    \caption{\label{windingscheme} $a)$ Representation of the winding process in the discrete setting, with  a layer of thickness $h$  is added to the  coil with a pulling force $F$ applied. $b)$ Continuous representation of the same process under the assumption that   $h \to 0$ and $F \to 0$, while the ratio $\sigma_a=F/h$ remains  finite. Winding starts off from  an internal rigid core of radius $R$, with   externally controlled  current placement of the growth surface   $\psi(t)$ (red line).}
\end{figure}

We assume for simplicity that the winding process is two-dimensional, taking place around a cylindrical  core.  The grown body is then a hollow disk that remains confined to a plane, see Fig.\ref{windingscheme}.  At the micro-scale,  winding can be described as sequential addition of pre-stressed thin strips of thickness $h$ to a growing core region.  A controlled tangential pulling force is applied on each individual segment (strip) as it adheres to the growing core. 

Denote by $F$   the tangential pulling force  and   suppose   that  in the   macroscopic approach  we are dealing with the limit  $h \to 0$ and  $F \to 0$, with  the ratio $\sigma_a=F/h$ remaining  finite, see Fig.\ref{windingscheme}.  The  task of developing an adequate macroscopic continuum model of winding is assisted by the fact that, in contrast to the case of solidification, we have access to a physically transparent micro-model operating with finite values of parameters $h$ and $F$.
 
For consistency with the  microscopic picture,  we assume that in the  macroscopic continuum formulation the body is growing radially while preserving cylindrical  symmetry.  We also assume for simplicity that   post-deposition slips between the neighboring strips are prohibited, allowing us to use the limit of ``frozen'' inelastic strain $\beps_p(\bx)$.  As another simplification, we also prescribe the  function $\vartheta(\bx)$ that describes the location of the deposition surface. To summarize, roll winding in our formulation is viewed as a process where the acquired incompatibility  is  controlled by the radial winding velocity $D(t)$ and by the tangential pulling force $\sigma_a(r,t)$. 

An important   goal  of our analysis  will be to compare two  solutions of the  formulated problem. One of them, is a purely macroscopic solution obtained by solving incremental equilibrium equations outlined in Sec.~\ref{casestudy} and by making no explicit reference to micro-displacements. Another one, is a solution of the full boundary value problem involving micro- and macro-displacements as formulated in Sec.~ \ref{dynamics1}. Such comparison is possible because our formulation of the winding problem is compatible with the decoupling of micro- and macro-evolutions.

 \subsection{Macroscopic approach} \label{Sec:incremental}

For additional analytical transparency we ignore space and time dependency of the controls and we assume that 
\beq D(t) \equiv D, 
\qquad
\sigma_a(r,t) \equiv \sigma_a
\eeq
where $D$ and $\sigma_a$ are given constants. In particular,  this means that  the growth radius evolves according to  $\psi(t) = R + D t$, where $R$ denotes the initial radius of the coil, interpreted as a rigid mandrel.
The function $t=\vartheta(r)$ representing the time when the deposition  surface is at the radial distance $r$, is obtained by inverting $\psi(t)$, giving (see Fig.\ref{windingscheme})  
\beq 
\vartheta(r)=(r-R)/D. 
\eeq 
Because winding takes place under null macroscopic loading, while  the microscopic pulling force is purely tangential, we conclude  that  on the deposition surface 
\beq
s_n(\psi(t)) = 0,\qquad\sigma_{\tau}(\psi(t),t)=\sigma_a.
\eeq
Note next that because loading is radially symmetric, the  incremental displacements are purely radial, 
\beq
\dot\bu(r,t)=\dot u(r,t)\be_r
\eeq 
where $\be_r$ is the radial unit vector and the superposed dot indicates differentiation with respect to time, $\dot{u} = \partial u/\partial t$. The resulting incremental strain is 
\beq
\dot\beps(r,t)=\dot\epsilon_r(r,t)\be_r\otimes\be_r+\dot\epsilon_{\theta}(r,t)\be_{\theta}\otimes\be_{\theta}
\eeq 
where $\be_{\theta}$ is the angular unit vector and we introduced notations   $\dot\epsilon_r=\dot u'$, and $\dot\epsilon_{\theta}=\dot u/r$, where superposed apex now indicates differentiation with respect to radial distance, $u' = \partial u/\partial r$.

The incremental Cauchy stress is then 
\beq
\dot\bsigma(r,t)=\dot\sigma_r(r,t)\be_r\otimes\be_r+\dot\sigma_{\theta}(r,t)\be_{\theta}\otimes\be_{\theta}
\eeq
 where $\sigma_{\theta}$ is the the hoop stress  representing in this setting   the surface stress  $\sigma_{\tau}$. Because according to our assumptions, inelastic strains are frozen at the moment of deposition, the incremental linear elastic constitutive relations take  the form 
\beq \label{Lame1}
\dot\sigma_{r/\theta}=2\mu\dot\epsilon_{r/\theta} + \lambda(\dot\epsilon_r+\dot\epsilon_{\theta}). 
\eeq
Note  that in \eqref{Lame1}   we used  the 2D Lamé coefficients $(\lambda,\mu)$, that are  related  to the corresponding 2D elastic modulus $E$ and the 2D Poisson coefficient $\nu$ through the relations $\mu=E/(2(1+\nu))$ and $\lambda=\nu E/(1-\nu^2)$.

 \subsubsection{Incremental  problem}

 As in the general case  discussed in Sec.\ref{casestudy} we can write  the incremental equilibrium equation in the form 
\beq \label{eqinc1}
\Div\dot\bsigma=\textbf{0}
\eeq
which in the presence of radial symmetry  reduces to  
\beq
\dot\sigma_r'+(\dot\sigma_r-\dot\sigma_{\theta})/r=0. 
\eeq
The  resulting mathematical problem for the unknown field $\dot u (r,t)$  takes the form  
\beq \label{general2}
\left\{
\begin{array}{lll}
\dot\e_r = \dot u'\\
\dot\e_{\theta}=\dot u/r\\
\dot\sigma_{r/\theta}=2\mu\dot\e_r + \lambda(\dot\e_r + \dot\e_{\theta})\\
\dot\sigma_r'+(\dot\sigma_r-\dot\sigma_{\theta})/r=0.
\end{array}
\right. 
\eeq
To obtain the boundary condition on the deposition surface we need, like we did more in general in \eqref{incrpb11}, to take the divergence of the surface stress 
     \beq
     \bbar\bsigma(\bx)=\bsigma(\bx,\vartheta(\bx))
       \eeq
which gives the condition  
   \beq \label{eqinc2}
\dot\bsigma\bn=D\,\Div\bbar\bsigma.
       \eeq
By specializing  \eqref{eqinc2} to our case, we obtain 
\beq
\dot\sigma_r(\psi(t),t) = - D \frac{\sigma_a}{\psi(t)}.
\eeq
We reiterate  that in our setting the parameters  $D$ and $\sigma_a$ as well as the function  $\psi(t)$ are   assumed to be known.  Recall also that in view of \eqref{extraconds1}, the radial displacement can be written in incremental form  
\beq\label{incradu}
\tilde u(r,t) = \int_{\vartheta(r)}^t\dot{u}(r,s)\,ds. 
\eeq
Finally, because on the  internal surface (the rigid core) we assume that $u(R,t)=0$ for all $t$, the corresponding  incremental displacement field must satisfy
\beq
\dot u(R,t) = 0. 
\eeq

 \subsubsection{Solution of the incremental  problem}

\begin{figure}[h!]
\centering
\includegraphics[width=\linewidth]{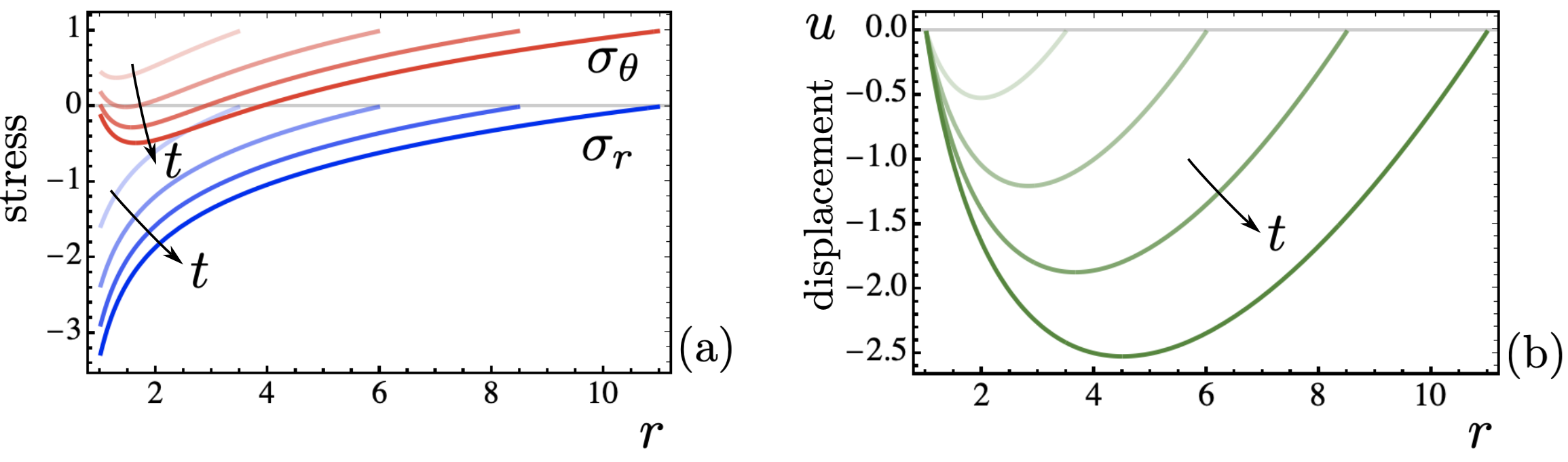}
\caption{\label{windingstrdis} (a) Distribution of radial ($\sigma_r$) and hoop ($\sigma_{\theta}$) stress components during the winding process  advancing according to $\psi(t) = R + D t$, with prescribed $D$. (b) Displacement field in the coil during deposition. Parameters: $\sigma_a = E = 1$, $R = 1$, $D = 1$, $\nu = 1/3$.}
\end{figure}

The  general solution of the problem \eqref{general2} is 
\beq
\dot u(r,t)=a(t)r+b(t)/r 
\eeq 
with the unknown functions  $a(t),b(t)$  to be obtained from the incremental boundary conditions
\beq
\dot u(R,t) = 0, \qquad \dot\sigma_r(\psi(t),t) = - D {\sigma_a}/{\psi(t)}. 
\eeq
After the incremental problem is solved, the final displacement field and the corresponding  stress distribution  can be recovered (as  in Sec.\ref{casestudy}) using \eqref{incradu} together with 
\beq
\sigma_{r/\theta}(r,t) = \bar\sigma_{r/\theta}(r) + \int_{\vartheta(r)}^t\dot \sigma_{r/\theta}(r,s)\,ds.
\eeq 
The resulting relations can be written explicitly 
\beq\label{utildes}
\tilde u(r,t) = \frac{\sigma_a}{2}\frac{(r^2 - R^2)\,(1 - \nu)}{r\,E} \log\left(\frac{r^2 (1 + \nu) + R^2 (1 - \nu)}{\psi(t)^2 (1 + \nu) + R^2 (1 - \nu)} \right)
\eeq
\beq\label{sigmardisk}
\sigma_r(r,t) =\frac{\sigma_a}{2}\left(1 + \frac{R^2 (1 - \nu)}{r^2\,(1 + \nu)} \right)\log\left(\frac{r^2 (1 + \nu) + R^2 (1 - \nu)}{\psi(t)^2 (1 + \nu) + R^2 (1 - \nu)} \right)
\eeq
\beq\label{sigmathdisk}
\sigma_{\theta}(r,t) =\sigma_a + \frac{\sigma_a}{2}\left(1 - \frac{R^2 (1 - \nu)}{r^2\,(1 + \nu)} \right)\log\left(\frac{r^2 (1 + \nu) + R^2 (1 - \nu)}{\psi(t)^2 (1 + \nu) + R^2 (1 - \nu)} \right).
\eeq
They  are illustrated in Fig.\ref{windingstrdis}.   In particular, panel (a) shows the distribution of radial and circumferential stress at successive time instants. As expected, the radial stress is always compressive and vanishes at the outer radius, while the circumferential stress is tensile and constant at the outer boundary. Inside the domain, it can be either tensile or compressive.
In panel (b), we show the displacement of material points  which are always negative as intuitively expected, because during winding the radial compression increases, leading to inward overall motion. 

The inelastic strains in the forming coil are obtained from 
\beq
\tilde\gamma_{r/\theta}(r)=\epsilon_{r/\theta}(r,t) - e_{r/\theta}(r,t)
\eeq
where the total strains are $\epsilon_r(r,t)=\tilde{u}'(r,t)$ and $\epsilon_{\theta}(r,t)=\tilde{u}(r,t)/r$ and where the components of the elastic strain are 
\beq
e_{r/\theta}=(\sigma_{r/\theta}(r,t)-\nu\sigma_{\theta/r}(r,t))/E.
\eeq
Using our explicit solution, we finally obtain the components of inelastic strain in the form
\beq\label{gammatildes}
\tilde\gamma_r(r) = \frac{\sigma_a}{E}\frac{r^2(1+\nu)-R^2(1-\nu)}{r^2(1+\nu)+R^2(1-\nu)}, \qquad
\tilde\gamma_{\theta}(r) = - \frac{\sigma_a}{E}.  
\eeq
As expected, the inelastic strains are time independent, and the resulting incompatibility carries  both regular and singular  contributions. The regular contribution $\boldsymbol{\eta}=\Curl\Curl\beps_p$, can be written in our case in the  form 
\beq
\boldsymbol{\eta}=\eta(r)\bk\otimes\bk, 
\eeq 
where
\beq\label{etareg}
\eta(r) = \tilde\gamma_r''(r)+\frac{1}{r}(2\tilde\gamma'_{\theta}(r)-\tilde\gamma'_{r}(r)) = \frac{4 R^2(1-\nu^2)\sigma_a}{E(r^2(1+\nu)+R^2(1-\nu))^2}. 
\eeq
The   singular  part appears because the  domain is non-simply connected.  It represents  a wedge disclination with the opening angle equal to 
\beq\label{etasing}
\Omega_p = 2\pi R\left(\tilde\gamma_{\theta}'(R)+\frac{\tilde\gamma'_{\theta}(R)-\tilde\gamma'_r(R)}{R}\right) = - 2\pi \frac{1+\nu}{E}\sigma_a
\eeq
see \cite{ZTPRL,TZPRE} for details, where a similar problem was considered. 

\subsubsection{Compatible component of inelastic strain\label{CompStress}}

It is important to stress that the final distribution of residual stress \eqref{sigmardisk}, \eqref{sigmathdisk},  emerging  at the end of the winding process, contains not only incompatible (regular and singular), but also compatible contribution which can be linked to the presence of an external constraint represented by the rigid mandrel. 

Indeed, we can show that even after a complete relaxation of the micro-displacement tensor $\Vtau$, as it would be the case if  we activate the relaxation process described by  \eqref{Pbulk2}  with $\B_a=\0$, the  grown coil will still  support nonzero residual  stresses of compatible nature. To determine their structure we need to solve the problem
\beq
\left\{
\begin{array}{lll}
\bsigma = 2\mu\beps_e + \lambda(\tr\beps_e)\I\\
\Div\bsigma = \0\\
\Curln\bsigma = \0
\end{array}
\right. 
\eeq
where $\beps_e$ is viewed as an unknown field. Under the assumption of  radial symmetry, the diagonal components of elastic strain  $(e_r, e_{\theta})$ and the corresponding components of stress $(\sigma_r, \sigma_{\theta})$ must satisfy the system of equations
\beq\label{stressrelaxgen}
\left\{
\begin{array}{lll}
\sigma_{r/\theta}=2\mu e_{r/\theta} + \lambda(e_r+e_{\theta})\\
\sigma_r'+(\sigma_r-\sigma_{\theta})/r = 0\\
\sigma_{\theta}'+(\sigma_{\theta}-\sigma_r)/r = 0. 
\end{array}
\right. 
\eeq
By straightforward  integration we find that 
\beq \label{epsilon0core11}
e_{r/\theta} = c_1 \pm \frac{c_2}{r^2}
\eeq
where $c_{1/2}$ are integration constants.  It is easy to check  that such elastic strain field carries neither regular nor singular  incompatibility. Their compatible origin is, of course, in line with our discussion in Sec.~\ref{relaxation1}.

To identify the compatible component in the  final distribution of residual stress \eqref{sigmardisk}, \eqref{sigmathdisk},  we can use linearity of the problem and consider two boundary value problems. In both problems we use the  inelastic strains $(\tilde\gamma_r,\tilde\gamma_{\theta})$ acquired  at the end of winding, see  Eq.\eqref{gammatildes}, as the internal loading. 

The first problem, $\mathcal{P}_1$, will have the  boundary conditions: 
\beq
\left\{
\begin{array}{lll}
\sigma_r(\psi_f) = 0\\
u(R) = 0.
\end{array}
\right. 
\eeq
Since the  solution of the ensuing problem describes the configuration of the grown coil  the resulting  stress distribution  coincides with the one represented by  \eqref{sigmardisk}, \eqref{sigmathdisk} with $\psi(t)=\psi_f$. 

The second problem, $\mathcal{P}_2$, is characterized by different set of boundary conditions:  
\beq
\left\{
\begin{array}{lll}
\sigma_r(\psi_f) = 0\\
\sigma_r(R) = 0
\end{array}
\right. 
\eeq
and describes the grown coil without the internal mandrel: in this case stresses are solely due to incompatibility. 

By solving these two linear problems and then subtracting the corresponding  stress distributions we can isolate the contribution to the plastic strains $(\tilde\gamma_1,\tilde\gamma_2)$ that arises from the rigid core alone.  We will refer to it as a solution of the difference problem $\mathcal{P}_0=\mathcal{P}_1-\mathcal{P}_2$.

\begin{figure}[h!]
    \centering
    \includegraphics[width=\linewidth]{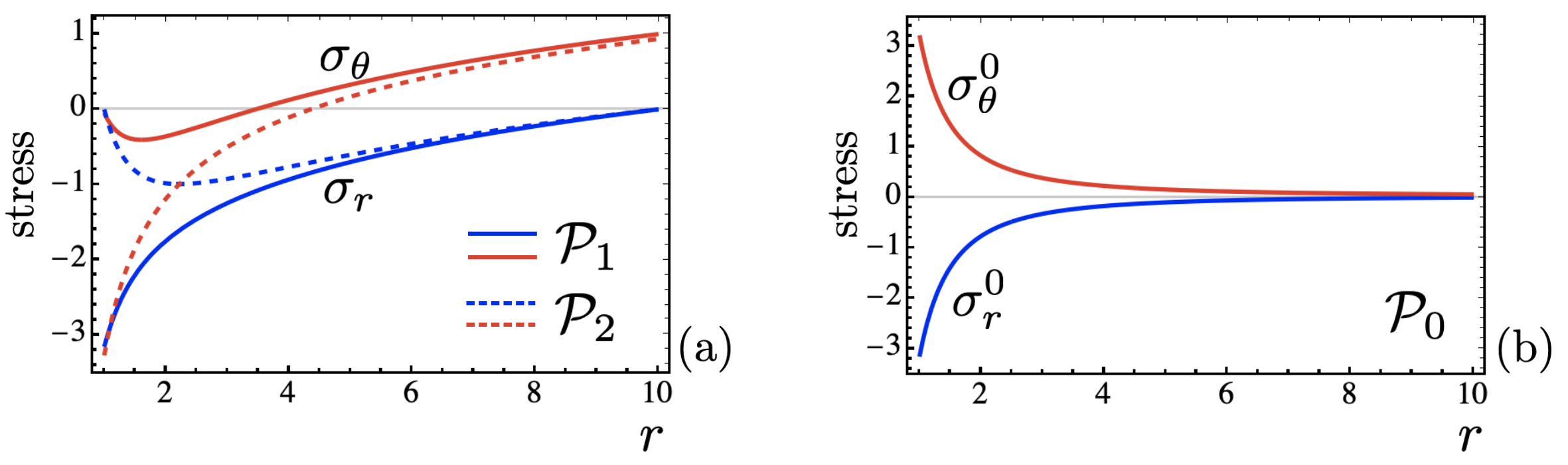}
    \caption{\label{core} (a-solid) Distribution of radial ($\sigma_r$) and hoop ($\sigma_{\theta}$) stresses at the end of winding process.  These residual stresses are due   both to incompatibility   and to compatible plastic strains originating from the presence of the rigid core (problem $\mathcal{P}_1$); (a-dotted) Distribution of radial ($\sigma_r$) and hoop ($\sigma_{\theta}$) stresses that are due solely to incompatibility (problem $\mathcal{P}_2$); (b) Distribution of radial ($\sigma_r$) and hoop ($\sigma_{\theta}$) stresses  in the difference problem $\mathcal{P}_0=\mathcal{P}_1-\mathcal{P}_2$. Parameters: $\sigma_a = E = 1$, $R = 1$, $D = 1$, $\nu = 1/3$.}
\end{figure}

After performing these operations we obtain the stress distribution  in the difference problem $\mathcal{P}_0$ in the form, 
\beq\label{sigma0core}
\sigma_{r/\theta}^0= \frac{R^2\sigma_a(\pm\psi_f^2-r^2)}{r^2(1+\nu)(R^2-\psi_f^2)}\log\left(\frac{R^2(1-\nu)+(1+\nu)\psi_f^2}{2R^2}\right).
\eeq
If we now compute the associated elastic strain field
\beq\label{epsilon0core}
e^0_{r/\theta}= \frac{\sigma^0_{r/\theta} - \nu\,\sigma^0_{\theta/r}}{E}
  = c_1^0 \pm \frac{c_2^0}{r^2}
\eeq
where the constants  $c^0_{1/2}$ are  determined by the boundary conditions. One can see that  we  obtained  the strain distributions  of the form  \eqref{epsilon0core11}, which implies  that such strains carry no incompatibility.  This result is, of course,  expected since the  addition of  the mandrel is equivalent to elastic loading  producing  purely compatible strains, see Fig.\ref{core}$_b$. 

We end this section with some  comments of physical origin explaining the somewhat counter intuitive presence of compatible residual stresses in the grown coil.
Recall that real  coils are usually produced using a single continuous tape spiraling onto itself.  In contrast, our model coil is produced by adding   circular layers.  The absence of the   continuity of the tape  does not produce appreciable differences with experimental observations  in the winding stage (see e.g. \cite{Zabaras95}), however, it affects  the nature of the subsequent stress relaxation stage. 

Thus, if the layers are interconnected, and if the friction is absent, global sliding can fully  relax all the stresses. Instead, in our model,  which  treats layers as independent, only the incompatible contribution to inelastic stresses can be  relaxed. Therefore, the contribution  to  residual stresses  due to the compatible part of inelastic strains will survive as we explain in  detail in  Sec. \ref{rel-winding}.

 \subsection{Micro-displacement-based approach}

We now turn to the description of the same winding protocol using our new general approach, which relies on extended kinematics and is based on the concept of a continuously generated micro-displacement tensor field. 
As before, we assume that the location of the deposition surface is  prescribed in the form $\psi(t)=R+D t$, which allows us to omit the  evolution equation  \eqref{eqD3}. Furthermore, to separate the deposition problem from the relaxation problem (discussed in detail in the the next section),  we assume that $\alpha = \infty$ in \eqref{Pbulk2}  so that the micro-displacement tensor $\U (\bx)$ is frozen at deposition. 

The outer surface is subjected to a prescribed surface micro-scale stress 
$\bsigma_a = \sigma_a \be_{\theta} \otimes \be_{\theta}$, 
where the winding tension $\sigma_a$ is  a given constant. We also keep the assumptions that  winding takes place in 2D and  that radial symmetry is maintained.  Finally, the  macro-displacements are again assumed to be of the form $\bu(\bx,t)=u(r,t)\be_r$.

 \subsubsection{Dynamical problem}

Given that the normal $\bn$ to the growth surface coincides with the unit radial vector, $\bn=\be_r$, the natural ansatz for the micro-displacement tensor is of the form
\beq
\U=\V_{\tau} \hodge\be_r.
\eeq 
Furthermore, since the tensor $\Vtau$ is   conjugate to  surface stress,  these two tensors have the same spectral form and we can write
\beq
\Vtau=v(r)\be_{\theta}\otimes\be_{\theta}
\eeq
where we took into account that the tensor $\Vtau$  is assumed to be ``frozen'' and therefore  time-independent. The associated inelastic  strain is then 
\beq
\beps_p(r) = -\sym(\Curl(v(r) \hodge\be_r ))=\gamma_r(r) \be_r\otimes\be_r + \gamma_{\theta}(r) \be_{\theta}\otimes\be_{\theta}
\eeq
where the radial and hoop components are obtained from $v(r)$ as
\beq\label{gamma-disk}
\gamma_r(r) = - \frac{v(r)}{r}, \qquad
\gamma_{\theta}(r) = - v'(r)
\eeq
where we recall that  $'=\partial_r$. The dependence  of the  total strain  on  macro-displacement must be of a ``complementary'' form and indeed as we already know 
\beq
\epsilon_r(r,t) = u'(r,t),\qquad\epsilon_{\theta}(r,t) = \frac{u(r,t)}{r}. 
\eeq
Radial and hoop components of the elastic strain can thus now written  in the form 
\beq\label{bepps}
e_r(r,t)  = u'(r,t) +\frac{v(r)}{r}, \qquad e_{\theta}(r,t)  = v_r'(r) +\frac{u(r,t)}{r}. 
\eeq
Note that in our previous incremental formulation, the micro-displacement structure of $e_r(r,t) $ and $e_{\theta}(r,t)$ had remained \emph{invisible}.

Radial symmetry also allows to represent the Cauchy stress in the form 
\beq
\bsigma = \sigma_r(r,t)\be_r\otimes\be_r + \sigma_{\vartheta}(r,t)\be_{\theta}\otimes\be_{\theta}
\eeq 
where $(\sigma_r,\sigma_{\vartheta})$ are the radial and hoop components of the stress, constitutively given by the relations 
\beq\label{constdisk}
\sigma_{r/\theta} =2\mu {\varepsilon_e}_{r/\theta} + \lambda({\varepsilon_e}_r+{\varepsilon_e}_{\theta}).
\eeq 
With $\bn=\be_r$ and $\P_r=\be_{\theta}\otimes\be_{\theta}$, the surface stress on  the growth surface can be written as 
\beq
\bbar\bsigma_{\tau} = \P_r\bbar{\bsigma}\P_r=\bbar{\sigma}_{\theta}\be_{\theta}\otimes\be_{\theta}
\eeq 
where, as before, we adopt the notation
$ \bbar{A}(r) = A(r, \vartheta(r))$ with $ \vartheta(r) = (r - R)/D $, to denote the value of a time-dependent field $A(r, t)$ at the instant when the growth surface is located at the radial position $r$.

At this stage, the unknown fields are  $u(r,t)$ and $v(r)$. The equilibrium equation \eqref{Pbulk1} takes the form
\beq\label{eqnswinding}
\sigma_r' + \dfrac{\sigma_r - \sigma_{\vartheta}}{r} = 0,\qquad R < r < \psi(t).
\eeq
To complete the  formulation of the problem we finally add boundary conditions. The ``macroscopic'' boundary conditions at the deposition surface $r=\psi(t)$ and at the internal core $r=R$ are  
\beq\label{bndu}
\left\{
\begin{array}{llll}
\sigma_r(\psi(t),t) = 0 \\
u(R,t) = 0.
\end{array}
\right.
\eeq
The corresponding ``microscopic'' boundary conditions  are  
\beq\label{bndv}
\left\{
\begin{array}{llll}
\sigma_{\theta}(\psi(t),t) = \sigma_a, \\[6pt]
v(R) = 0
\end{array}
\right.
\eeq
 After substitution of  \eqref{constdisk} into \eqref{eqnswinding}$_1$, the latter takes the form
\beq \label{w1}
w(r,t)-r(w'(r,t) + r w''(r,t))=0
\eeq
where we have introduced, for convenience, the function
\beq\label{eq-w}
w(r,t)=u(r,t)+\nu v(r).
\eeq
Note that in the incremental macro-scale formulation, the micro-displacement component of $w(r,t)$ had remained  invisible.

\subsubsection{Solution of the dynamical problem}

The  general solution  of  \eqref{w1} is 
  \beq
 w(r,t) = A(t)/r+B(t) r.
 \eeq
By imposing the boundary conditions \eqref{bndu} we obtain 
\beq\label{AAA}
A(t) = \frac{R (1 + \nu) \, \psi(t) \left[ R (-1 + \nu) \, v(\psi(t)) - \nu \, v(R) \, \psi(t) \right]}{R^2 (-1 + \nu) - (1 + \nu) \, \psi(t)^2}
\eeq
\beq\label{BBB}
B(t) = \frac{(-1 + \nu) \left[ R \nu \, v(R) - (1 + \nu) \, v(\psi(t)) \, \psi(t) \right]}{R^2 (-1 + \nu) - (1 + \nu) \, \psi(t)^2}.
\eeq
If we now substitute  $$u(r,t) = A(t)/r + B(t) r - \nu v(r)$$ into \eqref{bndv}$_1$ and   use the boundary condition \eqref{bndv}$_2$, we obtain a linear, first order ODE for $v(r)$ whose  solution is 
\beq\label{v-end-winding}
v(r) = \frac{\sigma_a}{2E} \left( r + \frac{R^2 (1 - \nu)}{r (1 + \nu)} \right)\cL(r).
\eeq
Here we have introduced  for further convenience  an  auxiliary function
\beq
\cL(a)=\log\left(\frac{R^2 (1 - \nu) + a^2 (1 + \nu)}{2 R^2} \right). 
\eeq
Using \eqref{v-end-winding} we  can also write the explicit expression for  macro-displacements 
\beq\label{umicro}
u(r,t) = -\frac{\sigma_a}{2 r E (1 + \nu)} \left[
\nu \left( R^2 (1 - \nu) + r^2 (1 + \nu) \right) \cL(r)
+ (r^2 - R^2)(1 - \nu^2) \cL(\psi(t))
\right].
\eeq
By substituting \eqref{umicro}  into \eqref{gamma-disk} we   obtain a  representation of the   inelastic strain in the form
\beq\label{gammarmicro}
\gamma_r(r) = -\frac{\sigma_a}{2 E}\frac{R^2 (1 - \nu) + r^2 (1 + \nu)}{r^2 (1 + \nu)} \cL(r)
\eeq
\beq\label{gammathetamicro}
\gamma_{\theta}(r) = -\frac{\sigma_a}{2 E} \left(2 + \frac{R^2 (1 - \nu) + r^2 (1 + \nu)}{r^2 (1 + \nu)}  \cL(r) \right). 
\eeq
To compare the obtained results with  the solution of the  macroscopic  winding problem we need to  compute the expressions for radial and hoop components of the stress $\bsigma(\bx)$, and  for the regular and singular components of incompatibility $\eta(\bx)$. A straightforward substitution shows the full agreements with the results obtained in the previous section by using incremental approach, see our formulas \eqref{sigmardisk},\eqref{sigmathdisk},\eqref{etareg},\eqref{etasing}. This  confirms the equivalence of the two formulations, at least at the level of the macroscopic fields $\sigma(\bx)$ and  $\eta(\bx)$. 

Note, however, that a formal comparison between the expressions of $\bu(\bx,t)$ and $\beps_p(\bx)$ obtained through the micro-displacement formulation (see Eqs.~\eqref{umicro}, \eqref{gammarmicro}, \eqref{gammathetamicro}), and those of $\tilde{\bu}(\bx,t)$ and $\tilde{\beps}_p(\bx)$ derived within the incremental framework (see Eqs.~\eqref{utildes}, \eqref{gammatildes}), reveals a discrepancy. However, the fact that $\bu \neq \tilde{\bu}$ and $\beps_p \neq \tilde{\beps}_p$ simply reflects the underlying reparametrization invariance, as $\beps_p$ and $\tilde{\beps}_p$ differ by a compatible deformation. Specifically we easily find that
\beq
\beps_p(\bx)-\tilde\beps_p(\bx) = \sym\nabla\g(\bx)
\eeq
where $\g(\bx)=g(r)\be_r$  and 
\beq
g(r) = \frac{R^2(\nu-1)+r^2(\nu+1)}{2 r E (1+\nu)}\sigma_a\log\left(\frac{2 R^2}{R^2(1-\nu)+r^2(1+\nu)}\right).
\eeq
As it follows from \eqref{umicro} and \eqref{utildes}  
\beq
g(r) = u(r,t) - \tilde u(r,t) 
\eeq
which reflects the fact that  the compatible component of inelastic strain  can be  always absorbed into  the macro-displacement field and ultimately viewed as a simple re-parametrization of the reference coordinates.

\subsubsection{Compatible component of micro-displacement field \label{Sec-comp-vc}}

As we have  shown in Sec.\ref{CompStress}, the  inelastic strains at the end of the winding process contain, in addition to  a component that is  responsible for incompatibility,  another contribution which is fully  compatible and  which cannot  not be affected by the postulated   relaxation mechanism. 

Since  the micro-displacement tensor serves as the source of the inelastic strain, we can now  identify the part of $v(r)$ which is  responsible for incompatibility and which can be then viewed as the   source of residual stresses in the final configuration after the rigid core is removed.

We begin by identifying the  compatible  component  of the micro-displacement field. Suppose that regular and singular sources of incompatibility are absent and therefore 
\beq
\eta = \gamma''_{\theta} + \frac{2\gamma'_{\theta} - \gamma'_r}{r} = -\frac{v-rv'+2r^2 v''+r^3 v'''}{r^3} = 0
\eeq
and 
\beq\label{VolterraDisk}
\Omega_p =
2\pi R\left(\gamma_{\theta}'(R)+\frac{\gamma'_{\theta}(R)-\gamma'_r(R)}{R}\right) 
= 2\pi \left(\frac{v(R)}{R}-v'(R)-R v''(R)\right) = 0. 
\eeq
Solving these equations we conclude that  a compatible micro-displacement field is of the general form
\beq\label{fullycompv}
v_c(r) = {c_1}/{r} + c_2 r.
\eeq
As expected, the resulting plastic strain components are precisely of the compatible type already discussed in Sec.\ref{CompStress}. Specifically, given that  $$\bbeta_{cp}=({\gamma_c}_r ,{\gamma_c}_{\theta})=-\Curl(\Vtau\hodge\be_r)$$ with $$\Vtau=v_c\P_{\be_r},$$ we obtain that
\beq
{\gamma_c}_r = - \frac{v_c}{r} = - c_2 - \frac{c_1}{r}
\qquad
{\gamma_c}_{\theta} = - v_c' = - c_2 + \frac{c_1}{r}.
\eeq
The compatibility means that one can find the  corresponding macro-displacement field
\beq \label{fullycompv1}
\bu(\bx)=(c_1/r-c_2 r)\be_r
\eeq
which has the property $\sym\nabla\bu\equiv\sym(\Curl\U)$ with $\U=\Vtau\hodge\be_r$. In other words,  the compatible component of the micro-displacement field $v_c(r)$  can be equivalently represented by a macro-displacement field $\bu$. 

To find explicitly the compatible field $v_c(r)$ which is relevant for our problem, we need to fix the constants $c_1,c_2$ by imposing the conditions 
\beq\label{hombndvc}
v_c(R)=v(R), \qquad v_c(\psi_f) = v(\psi_f)
\eeq
where $v(r)$ is the micro-displacement field at the end of winding \eqref{v-end-winding}. We obtain
\beq\label{c1c2}
c_1 = \frac{R\psi_f(R v(\psi_f) - \psi_f v(R))}{R^2-\psi_f^2}, \qquad
c_2 = \frac{R v(R) - \psi_f v(\psi_f)}{R^2-\psi_f^2}. 
\eeq
At this point, the incompatible component of the micro-displacement field 
\beq\label{vinc-def}
v_i(r):=v(r)-v_c(r)
\eeq
can be now found explicitly as
\beq\label{vi-final}
v_i(r) = \frac{\sigma_a}{2E} \left( q_1(r)\, q_2(r) - q_4(r)\, q_3 \right),
\eeq
where we have set  
\beq
\left\{
\begin{array}{lll}
\displaystyle q_1(r) = r + \frac{R^2(1 - \nu)}{r(1 + \nu)}, \\
\displaystyle q_2(r) = \log\left( \frac{1}{2} \left( 1 - \nu + \frac{(1 + \nu) r^2}{R^2} \right) \right), \\
\displaystyle q_3 = \log\left( \frac{1}{2} \left( 1 - \nu + \frac{(1 + \nu) \psi^2}{R^2} \right) \right), \\
\displaystyle q_4(r) = \frac{(r^2 - R^2)\left[ R^2(1 - \nu) - (1 + \nu)\psi^2 \right]}{r(1 + \nu)(R^2 - \psi^2)}.
\end{array}
\right. 
\eeq
Note, that in view of  \eqref{hombndvc}, the field $v_i$ automatically satisfies the boundary conditions $v_i(R)=v_i(\psi_f)=0$. Having isolated the  incompatible component of the micro-displacement field, we will use the distribution \eqref{vi-final} as the initial condition in the   relaxation process studied in Sec. \ref{rel-winding}.

\subsection{Discussion}

In this section we have presented a detailed case study of the process of  coil winding   under the simplifying assumptions  that the rate of material deposition is externally controlled  and the attachment-induced micro-displacement field  is ``frozen'', in the sense that there is no  post-deposition slip between the deposited layers. The obtained solutions reveal   that the micro-displacement tensor  field  $\U(\bx)$  necessarily contains compatible  and non-compatible additive components.  The analysis has shown that  the compatible component of $\U(\bx)$ is due to the rigid constraint provided by the rigid core, and that it relaxes elastically if such core is removed. Instead, the incompatible part of $\U(\bx)$ persists even after the removal of  the mandrel.

In our simplified  setting we were able to demonstrate   that  the macroscopic problem of finding the vector field $\bu(\bx,t)$ can be decoupled from the microscopic problem  of finding   the micro-displacement tensor  field $\U(\bx)$. We have also shown that the results obtained by solving first the macroscopic problem and then reconstructing the micro-displacement field are perfectly equivalent to the solution of our dynamical problem involving  simultaneously the fields $\bu(\bx,t)$  and  $\U(\bx)$. 

Although in this particular example both approaches could be used interchangeably, such an equivalence does not hold in general. Thus, in most realistic cases the function $\vartheta(\bx)$ cannot be prescribed a priori and  must be computed self-consistently. Similarly,  in general the  inelastic strains evolve together with the advancement of the growth surface, therefore these are not  ``frozen.''  For instance, as we have already seen in the solidification example discussed in the previous section, the problems for the macro- and micro-displacements could not be decoupled. In such situations, the proposed enriched kinematic framework becomes essential, as an adequate description of the growth process must necessarily involve both $\bu(\bx,t)$ and $\U(\bx,t)$. The success of such a coupled study is contingent upon the  proper treatment of the energetics of deposition.

\section{Bulk relaxation of inelastic strain}

In the two special cases solved in Sec.\ref{Sec-solid} and Sec.\ref{section-winding} we have assumed that no bulk relaxation of the inelastic  strain $\beps_p $ was allowed after the moment of deposition, meaning that we assumed $\alpha= \infty$ in \eqref{Pbulk2}. Here  we present a   study of  a  model problem  where  $\alpha < \infty $ and consider evolution governed by \eqref{Pbulk2}. To highlight the relaxational aspect of this example   we now assume for simplicity that the deposition process is over and the external boundary remains fixed in its  final  position  at $t=t_f$.

\subsection{Relaxation after solidification}

Referring to the problem discussed in Sec.~\ref{Sec-solid}  while maintaining, in particular,  the same horizontal stack geometry, we now assume that at the end of the   solidification process the distribution of micro-displacement $v(z)$ is known. The corresponding inelastic strain  emerges as an outcome of the deposition stage, and serves as the sole source of residual pre-stress in the grown body which is   revealed as soon as  the external  loads are removed. 

We recall that  the function $v(z)$   is obtained by integrating Eq.~\eqref{dv}  with the condition $v(0) = 0$ while also  using Eq.~\eqref{cF0}.  We obtain 
\beq \label{general1}
v^0(z) = \frac{1-\nu}{E}{P}({p_w}-{p_a})z \log\left(\frac{z}{\psi_0}\right). 
\eeq
The superscript $0$  reminds  that the corresponding distribution of the micro-displacement is now considered as the initial datum for the relaxation problem. To simplify the  analysis of the relaxation stage we neglect the nucleation layer  and we consider the unloaded configuration with $s_n=s_w=0$, defined within the segment $(\psi_0, \psi_f)$. Note that the dependency of $v^0(z)$ on $p_a,p_w$ is relative to the memory of the loading history, and not to the current loading condition during relaxation. 

We assume that during relaxation the system is mechanically equilibrated. This means that in the axial direction we have  
\beq \label{av-relax1}
\sigma_z'(z,t)= 0, \qquad (\psi_0 < z < \psi_f)
\eeq
while in the horizontal only the averaged equilibrium condition holds
\beq\label{av-relax}
\int_{\psi_0}^{\psi_f}\sigma_{\tau}(z,t)\,dz = 0. 
\eeq
Using \eqref{av-relax1} and the fact that the system is externally unloaded  we conclude  that    $\sigma_z=0$  and therefore 
\beq\label{u-relax}
u'(z,t) = \frac{2\nu}{\nu-1}(\epsilon_{\tau}(t) + v'(z,t)).
\eeq
If we now substitute  \eqref{u-relax}  into the averaged equilibrium equation \eqref{av-relax}, we obtain  
\beq\label{etau0}
\epsilon_{\tau}(t) = - \frac{v(\psi_f,t) - v(\psi_0,t)}{\psi_f-\psi_0}, 
\qquad
\sigma_{\tau}(z,t) = \frac{E}{1-\nu}(\epsilon_{\tau}(t) + v'(z,t)). 
\eeq
Using the obtained information we can rewrite  the evolution equation
\beq
\sigma_{\tau}'(z,t) = \alpha\dot v(z,t) \qquad (\psi_0 < z < \psi_f)
\eeq
in the form
\beq\label{evo-v-stack}
\frac{E}{1-\nu}v''(z,t) = \alpha\dot v(z,t) \qquad (\psi_0 < z < \psi_f). 
\eeq
It will be convenient to remove from the function $v(z,t)$ an additive part which automatically satisfies \eqref{evo-v-stack}, which specifically describes compatible micro- displacements. 

\begin{figure}[h!]
    \centering
    \includegraphics[width=\linewidth]{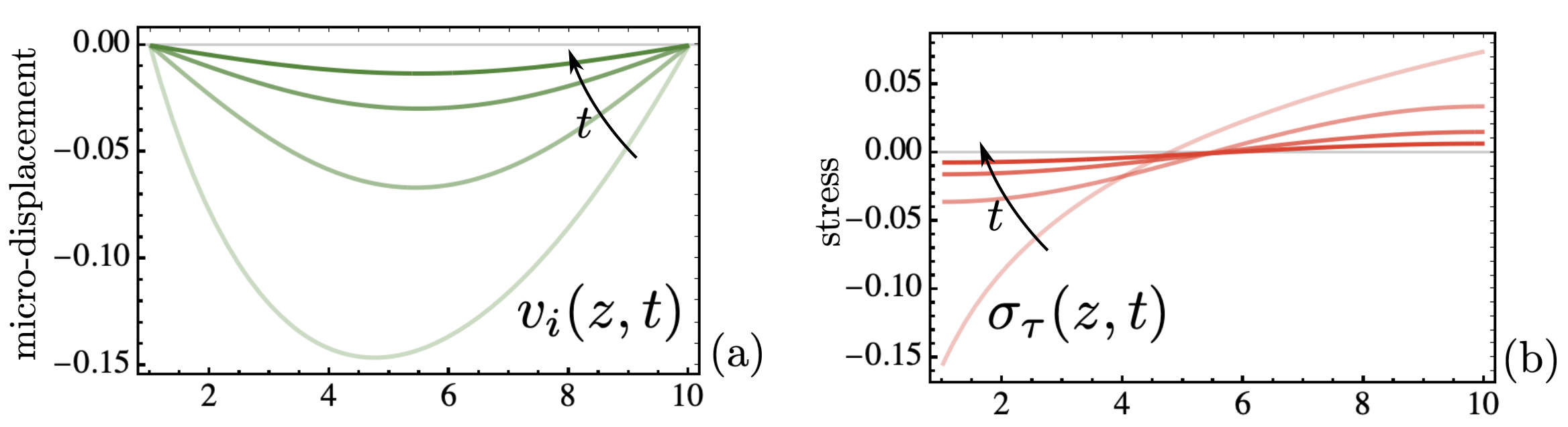}
    \caption{\label{relax-solid} Evolution of the horizontal stress (a) and incompatible micro-displacement (b) during the stress relaxation stage. Darker colors indicate more recent time instants. Parameters: $\sigma_a = E = 1$, $R = 1$, $D = 1$, $\nu = 1/3$, $\lambda=10$.}
\end{figure}

Indeed, we recall that according to \eqref{evo-v-stack} a linear contribution $v(z,t)=A(t) z + B(t)$ would not evolve  and,  as we have already already seen  in Sec.\ref{Sec-solid},  would  not   affect the incompatibility. Therefore we can define the compatible micro-displacement  field by the formula  $v_c(z)=A z + B$.  The time-independent constants $A,B$ can be found by imposing, like we did in \eqref{hombndvc} for the winding problem, that
\beq
v_c(\psi_0) = v^0(\psi_0), \qquad
v_c(\psi_f) = v^0(\psi_f)
\eeq
where $v^0(z)$ is given by \eqref{general1}. We thus obtain
\beq\label{vc-fixed-slab}
v_c(z) =\frac{{P}({p_w}-{p_a})(1-\nu)(z-\psi_0)\psi_f}{E(\psi_0-\psi_f)}\log\left(\frac{\psi_f}{\psi_0}\right). 
\eeq
The incompatible component of the micro-displacement, which relaxes according to \eqref{evo-v-stack}, may can be computed as
\beq
v_i(z,t) = v(z,t)-v_c(z)
\eeq
where by definition $v_i(\psi_0,t) = v_i(\psi_f,t)=0$, and where the initial value is 
\beq
v_i^0(z) = v^0(z)-v_c(z). 
\eeq
The ensuing evolution problem for $v_i(z,t) $ takes the form 
\beq\label{relax-solidification}
\left\{
\begin{array}{lll}
v_i''(z,t) = \omega^2 \dot v(z,t) \qquad (\psi_0<z<\psi_f)\\
v_i(z,0) = v_i^0(z)\\
v_i(\psi_0,t) = v_i(\psi_f,t) = 0
\end{array}
\right. 
\eeq
where we introduced the notation $\omega^2=\alpha(1-\nu)/E$.  The  problem \eqref{relax-solidification} is formulated for a linear  diffusion  equation which can be  solved explicitly, with results illustrated  in Fig. \ref{relax-solid}.  

According to Fig. \ref{relax-solid},  as time increases, the magnitude of incompatible micro-displacement $v_i(z,t)=v(z,t)=v_c(z)$
decreases throughout the whole domain.  This fully  relaxes residual  elastic stresses at the end of the process. 

The only component of the micro-displacement that survives  at the end of the relaxation process  is the one describing compatible  deformation and defined by \eqref{vc-fixed-slab}. According to  \eqref{etau0}$_1$,  where we set  $v(r,\infty)=v_c(r)$, this component  is responsible for a residual lateral strain 
\beq\label{epstau-relaxed}
\epsilon_{\tau}=\frac{{P}({p_w}-{p_a})(1-\nu)\psi_f}{E(\psi_0-\psi_f)}\log\left(\frac{\psi_f}{\psi_0}\right). 
\eeq
Similarly  from \eqref{etau0}$_2$ we infer  that at $t=\infty$ the lateral stress $\sigma_{\tau}$ will fully relax.

\subsection{Relaxation after winding\label{rel-winding}}

Consider next our second example where the pre-stressed configuration was obtained through winding a coil. We now assume that winding is complete and that the external boundary remains fixed in its final position at $t=t_f$, and starting from this final state we allow the incompatibility to relax according to \eqref{Pbulk2}. One of our goals will be to show that, once again, only the incompatible part of the micro-displacement field is involved in such relaxation process. 

Specifically,  we assume that  a  grown  coil  of radius $\psi_f$   carries the initial   micro-displacement distribution of the form
\beq\label{v-initial}
v(r,0) = v_0(r) \equiv \frac{\sigma_a}{2E} \left( r + \frac{R^2 (1 - \nu)}{r (1 + \nu)} \right) \log\left(\frac{R^2 (1 - \nu) + r^2 (1 + \nu)}{2 R^2} \right).
\eeq
Under the assumption of radial symmetry the governing  equations \eqref{Pbulk1}–\eqref{Pbulk2}  can be presented in the form of a system:
\beq\label{rad}
R < r < \psi_f: \quad
\left\{
\begin{array}{lll}
\sigma_r' + \dfrac{\sigma_r - \sigma_{\vartheta}}{r} = 0\\ [8pt]
\sigma_{\theta}' + \dfrac{\sigma_{\theta} - \sigma_r}{r} = \alpha\, \dot v
\end{array}
\right. 
\eeq
During relaxation, once again we find that the equilibrium equation \eqref{rad}$_1$ is satisfied as long as 
\beq
u(r,t) = A(t)/r + B(t) r - \nu v(r,t)
\eeq
where differently than in \eqref{eq-w}, the function $v(r,t)$ now depends also on time. The  functions  $A(t),B(t)$ are obtained by imposing the boundary conditions 
\beq\label{bndu-again}
\left\{
\begin{array}{llll}
\sigma_r(\psi_f,t) = 0\\
u(R,t) = 0. 
\end{array}
\right.
\eeq
For the coefficients $A(t),B(t)$ we naturally recover  the  same expressions \eqref{AAA}, \eqref{BBB} as in the previous section. 

Upon substitution of the obtain relations into \eqref{rad}$_2$, we end up with a linear PDE with variable coefficients, describing the relaxation of the micro-displacement field
\beq\label{Pbulk2s}
\alpha \dot v(r,t)=-\frac{E}{r}\left(\frac{v(r,t)}{r} - v'(r,t) - r v''(r,t)\right).
\eeq
Once again, if the micro-displacement field is compatible, $v(r,t)\equiv v_c(r,t)$ where, from \eqref{fullycompv}, 
\beq
v_c(r,t) = c_1 /r + c_2  r
\eeq
with $c_{1/2}(t)$ arbitrary functions of time, then equation \eqref{Pbulk2s} reduces to 
\beq
 \dot v_c(r,t)=0
\eeq
suggesting that for $v_c(r,t)$ to be a solution of \eqref{Pbulk2s}, the coefficients $c_{1/2}$ not depend on time, and therefore such micro-displacement field should be ``frozen'',  
\beq
v_c(r,t)=v_c(r). 
\eeq
Because the compatible component of   the micro-displacement field  $v(r,t)$ cannot change,   only  its   incompatible component   defined in \eqref{vinc-def} will be evolving   during the relaxation process described by \eqref{rad}$_2$. In other words,   the relaxation of the incompatible part of inelastic strain can be decoupled from the ``frozen''  compatible part.  The latter  can be, of course,  relaxed  elastically by removing the mandrel  without engaging a dissipative process.
 \begin{figure}[h!]
    \centering
    \includegraphics[width=\linewidth]{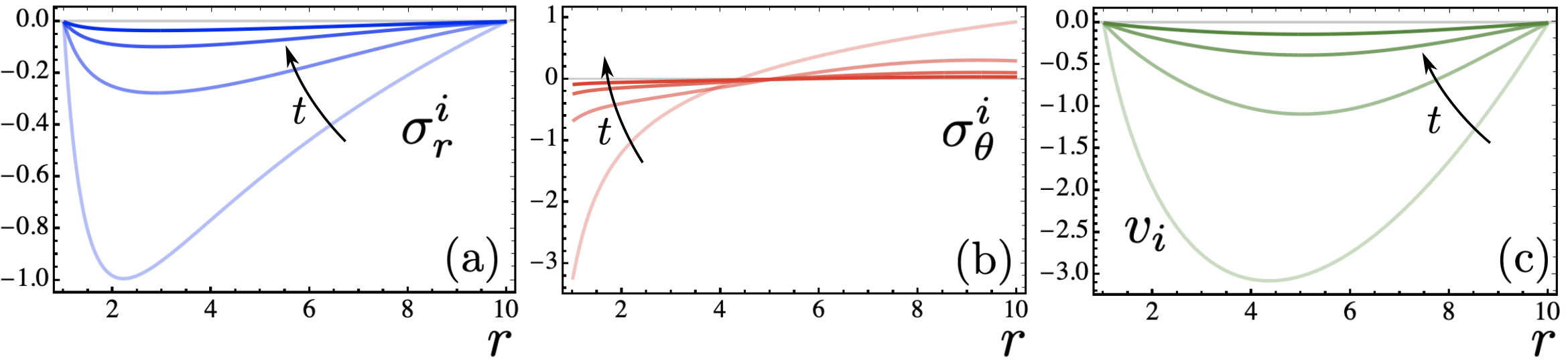}
    \caption{\label{evofinal} Evolution of the radial (a) and hoop (b) stress components, along with micro-displacement (c) during the stress relaxation stage. Darker colors indicate more recent time instants. Parameters: $\sigma_a = E = 1$, $R = 1$, $D = 1$, $\nu = 1/3$, $\lambda=10$.}
\end{figure}

To summarize, similar to what we have seen in the case of post solidification relaxation,  here again  the relaxation process described by \eqref{Pbulk2}  only concerns the  function
\beq
v_i(r,t): =v(r,t)-v_c(r)
\eeq 
where the compatible component $v_c(r)$ was determined in Sec.\ref{Sec-comp-vc}. The corresponding initial conditions are  
known since we  have  the expression for   $v_0(r)$ from \eqref{v-initial} and can write 
\beq
v_i(r,0):=v_0(r)-v_c(r). 
\eeq 
Since by definition 
\beq
v_i(R,0) = v_i(\psi_f,0) =0
\eeq
we assume that the same  boundary conditions hold throughout the whole relaxation process, so that 
\beq
v_i(R,t) = v_i(\psi_f,t) =0. 
\eeq
As we now focus on the evolution of $v_i(r,t)$ with boundary conditions $v_i(R,t) = v_i(\psi_f,t) =0$, we set $A=B=0$.  The associated  macro-displacements can be found from  $u(r,t) = - \nu v_i(r,t)$. 

The equations   governing the evolution of the incompatible micro-displacement field take the form
\beqn\label{vi-evo}
\left\{
\begin{array}{lll}
 \alpha\dot{v}_i(r,t) = -E\left({v_i(r,t)}/{r} - v_i'(r,t) - r v_i''(r,t)\right)/r \\
v_i(r,0) = v_i^0(r) \\
v_i(R,t) = 0 \\
v_i(\psi_f,t) = 0. 
\end{array}
\right. 
\eeqn
The problem  \eqref{vi-evo}$_1$ involves  a linear parabolic partial differential equation, of basically the same type as the equation obtained in the study of relaxation of solidification stresses, see  \eqref{relax-solidification}. The right-hand side of \eqref{vi-evo}$_1$ involves the radial part of the Laplacian acting on the incompatible micro-displacement field $v_i(r,t) $, while the left-hand side introduces a viscous/relaxation time scale proportional to $\alpha$. From the solution to \eqref{vi-evo} we can find the incompatible elastic strains  
 \beq
\epsilon^i_r = u' + v_i/r,\qquad   \epsilon^i_{\theta} = v_i' + u/r 
\eeq
and we can compute the associated components of stress, as exclusively induced by  incompatibility  
\beq
\sigma^i_r(r,t) = E v_i(r,t)/r,\qquad \sigma^i_{\theta}(r,t) = E v_i'(r,t). 
\eeq
The numerical solution to the problem \eqref{vi-evo} is illustrated in  Fig. \ref{evofinal}. As expected, the incompatibility-related  part of stress and the associated micro-displacement both  relax to zero. By the end of relaxation, the coil remains residually stressed due to the non-relaxing contribution $\sigma^0_{r/\theta}(r)$ from  \eqref{sigma0core} which, as we have seen, can be eliminated only  through the removal of the rigid core.

We emphasize once again that the persistence of residual stress after relaxation stems from the way the winding problem was  formulated. In particular, by  neglected the continuity of the   tape and  treating the winding process   as  sequential deposition of independent layers, we have eliminated 
the possibility for global ``unwinding''  relaxation modes. While the simplifications introduced here are justified by the resulting analytical transparency, further investigations are required for cases in which interlayer friction is overcome, thereby allowing the ensuing global relaxation process to completely eliminate the residual stresses within the coil \cite{Andreotti25}.

\subsection{Discussion}

In this section we have presented two simple models of bulk relaxation of inelastic strains accumulate during the process of surface deposition. We have assumed that relaxation is of a purely viscous (Onsagerian) type, while it could also be of a threshold type, as in dry friction and crystal plasticity. In this sense the adopted model can be viewed as representing a form of linear viscoelasticity.  

Our analysis reveals a fundamental decoupling between the compatible and incompatible components of the micro-displacement field: only the incompatible part evolves in time, while the compatible component remains “frozen” in its initial state. This indicates that the relaxation process in the adopted model is governed solely by incompatibility, which undergoes a dissipative evolution until the micro-displacement tensor becomes fully compatible. The resulting configuration thus preserves the residual stresses associated with the displacement-controlled boundary, as these cannot relax through the considered dissipative mechanism.

\section{Conclusions}

The  goal of  this paper is to  develop  an extended kinematics of continuum elastic bodies   allowing one to address micro-mechanical interactions  involved in the accumulation and storage of mechanical ``information'' inside a solid body.  The main assumption is  that such ``information'' is   carried by geometrical \emph{incompatibility}, which can be revealed, for instance,  by cutting the unstressed body into smaller sub-bodies while recording the resulting deformation. 

We are therefore concerned with the  question of  how the incompatibility can be \emph{inserted} into a solid body, and how the acquired mechanical  ``information'' may be  \emph{erased}  by  internal dissipative relaxation.

 The growing attention to such issues is mainly due to the emergence of novel applications of solid mechanics in the fields of biomechanics and bio-mimetic technology, where the task of microscopically supervised storage and maintenance of the incompatibility has recently come to the forefront, see  \cite{Renzi-ejsma-24,ZTPRL,ZTMRC,TZPRE,AEGZ24,AEGZ25} and the literature cited therein. It was found, for instance, that biological systems can develop incompatibility  by using metabolically supported  mechanical pathways, ultimately representing microscopic mechanical activity \cite{Farhadifar2007}.   
 Viewing incompatibility as being {\it managed} through internal activity, we have focused in this work on a specific case of controlled surface growth in non-Euclidean solids, where the role of the active microscopic mechanical “agents” responsible for inserting mechanical “information” can, for instance, be played by a 3D printer \cite{Zaza-ejmsa-21}.
 
To shed light on the  basic interplay between macro- and micro- mechanics in the process of active manipulation of  incompatibility, we  developed in this paper a novel kinematic approach in the general framework of continuum mechanics which explicitly identifies microscopic variables that are potentially acted upon by such microscopic active ``agents''. 

Our approach is built around the concept of micro-displacement \emph{tensor}, which complements the conventional kinematic description of solids based on the idea of a macro-displacement \emph{vector}. Given that  the physical dimensionality of   macro-displacement vector and micro-displacement tensor is  the same, in both cases the corresponding conjugate variables can be identified as  forces. However, if in the conventional macroscopic theory such  forces are represented by \emph{vectors}, their microscopic counterparts emerge as couples, represented by \emph{tensors}. We  have shown  that the proposed approach allows one to model the  externally imposed micro-mechanical controls at the boundary of the growing body  which go beyond the  conventional three components of surface tractions. Instead, rather counter-intuitively, they are adequately represented  by prescribing on the boundary of all six components of the macroscopic stress tensor. 

The idea that in a growing body, active micro stresses can effectively control  the emerging incompatibility was (even if implicitly) present in several  previous studies  \cite{Trincher, KingFletcher, Zabaras95, Rashba, Southwell, BrownGoodman}. However, while some of these  earlier treatments have already  imposed  a   phenomenological  assumption that the total state of stress   is prescribed  on the growth surface, they  did not operate with the  concept  of micro-displacements and therefore could  not adequately account for the energetic cost of the implied microscopic  activity. Also, in most of these works, the motion of the deposition surface was not treated self-consistently, as it was either prescribed from outside, or it was attributed to a  largely decoupled physical process, such as the removal of the released heat.   Instead, we showed that micro-forces may play a significant role in  mechanically regulated kinetics of deposition, by even  \emph{deciding} the  outcome of its long term evolution.
 
As there are still many challenges on the way of rigorous up-scaling of the newly defined microscopic quantities directly from the microscale, our approach to the \emph{energetics} of surface deposition is purely phenomenological. Specifically, we have developed our extended continuum mechanical description of associated active micro-mechanical processes with the reference to a simple layering \emph{ansatz} for the micro-displacement tensor, and by using only the most basic symmetry considerations and  thermodynamic inequalities. 

Using only classical thermodynamic considerations, we succeeded in deriving the equations governing both the process of incompatibility acquisition at the surface of the growing body, and the subsequent evolution of incompatibility past the initial state of deposition. In physical terms, our main assumption was that incompatibility can be generated by generalized micro-forces operating during deposition on the growth surface, and that its quasi-viscous relaxation can be in principle controlled by the action of  the generalized micro-forces, operating inside the growing body. While the former are exemplified by the agency responsible for the pre-stretch of the attached elastic bands, the latter can be pictured, for instance, as frictional forces preventing slips between adjacent deposition layers. 

We have illustrated our general constructions by three detailed case studies. The goal was to show how the proposed \emph{general} approach works in detail in \emph{special} situations, where geometrical and physical simplicity of the setting makes the problem amenable to an analytically transparent treatment.

Our first example addresses the classical solidification phenomenon, where  we assumed that the deposited solid material can  be internally pre-stressed in the process of  liquid-to-solid transition. While most previous  studies of  solidification had framed the problem within the ``elastic growth'' setting, with  solid phase growing incompatibility-free  \cite{HobbsOrd, Sekerka-Cahn-2004, GraTru23},  the  introduction of  a richer kinematics has allowed us to  extend the description to the case of  ``inelastic growth'', where the account of the work of microscopically generated active surface stresses  underlines their crucial role in the acquisition of incompatibility. 

Our second example can be interpreted as a minimal model of a winding process around a rigid mandrel, developed under the simplifying assumptions that the rate of material deposition is externally controlled and that the attachment-induced inelastic strain is “frozen” immediately after deposition. In this setting, we demonstrated that the macroscopic problem of determining the classical displacement field and the microscopic problem of determining the incompatibility tensor can be decoupled. We further showed that the pre-stretch of the deposited layers may give rise not only to incompatibility but also to compatible inelastic strain.

Our final example addressed the bulk relaxation of preexisting incompatibility, showing that only part of the growth-induced pre-stress can relax through the proposed viscous mechanism. Specifically, the analysis revealed a fundamental decoupling between the compatible and incompatible components of the micro-displacement field: only the incompatible part evolves under the adopted mechanism, while the compatible component remains “frozen” in its initial state.

 It is important to mention  that while in our three case studies  we considered \emph{separately} such  aspects of the surface deposition problem as the possibility of mechanically regulated self-consistent growth,  the unavoidable emergence of compatible inelastic strain   and  the  peculiarity  of mechanically controlled bulk relaxation of inelastic strain, in general applications one can expect that all these effects will appear \emph{together} and that only their delicate interplay will ultimately decide the mechanical outcome of the surface deposition process.    

The proposed model can be extended and enriched along several  directions. An important  first step could  be the development of a finite strain version of the same approach, e.g. \cite{TZPRE}.  A microscopically informed derivation of such nonlinear  model should   allow one  to establish a formal link of the proposed approach  with  the theory of second-order structured deformations, e.g. \cite{MatiasMorandottiOwen2023}.  However, the corresponding extended theory should not be restricted by the layering ansatz and should    also  account for the possibility of non-layering arrangement of the deposited material.
For instance,  while  considerable simplification was achieved in this paper due to the  simplifying assumption  that the  field  prescribing the orientation of the layering in the bulk is effectively ``frozen'' and  does not evolve past deposition, such an assumption is clearly not sufficiently general.  Similarly, our assumption regarding the  Onsagerian nature of relaxation can be   questioned, for instance,  one may argue that micro-displacements should relax in non-Onsagerian way as, for instance,  in   dry friction and plasticity. Furthermore, a more intricate coupling between micro and macro displacements can be imagined, for instance, in various biological applications one can expect that it is not even reciprocal \cite{lorenzana2025nonreciprocity}.  

An important step towards applications could be an adaptation of the proposed theory to particular technologically relevant modalities of 3D printing, including a specification of the mechanical micro-activity defining the writing head, e.g.  \cite{Zaza-ejmsa-21}. Another  challenge is  the adaptation of the developed  general theory to specific problems involving the manipulation of incompatibility   in living systems, where the abstract active ``agents'' would have to be identified  with  the  specific molecular micro-machines exposed to  metabolic energy sources, e.g  \cite{Julicher2007}.

\section{Acknowledgments}

The authors thank B. Shoykhet and P. Recho for helpful discussions. L.T.  acknowledges the support   under the grants ANR-17-CE08-0047-02,  ANR-21-CE08-MESOCRYSP an 2020-MSCA-RISE-2020-101008140. G.Z. gratefully acknowledges the kind hospitality extended by Politecnico di Bari while hosting his sabbatical leave from January to June 2025, where part of this research was conducted.

\bibliographystyle{elsarticle-num} 
\bibliography{bibtex.bib}

\end{document}